\documentclass[aps,pra,twocolumn,showpacs,superscriptaddress]{revtex4-1}  
\usepackage{graphicx}  
\usepackage{dcolumn}   
\usepackage{ogonek}    
\usepackage{bm}        
\usepackage{amssymb}   
\usepackage[fleqn]{amsmath}
\usepackage{amsthm}
\usepackage{bbold} 
\usepackage{mathrsfs}
\usepackage{array}
\usepackage[dvipsnames]{xcolor}
\usepackage{soul} 
\usepackage[normalem]{ulem} 

\begin{document}

\title{A mean-field approach to Kondo--attractive-Hubbard model}

\author{Natanael C. Costa}
\email{natanael@if.ufrj.br}
\email{natanael.c.costa@gmail.com}
\affiliation{Instituto de F\'\i sica, Universidade Federal do Rio de Janeiro, Rio de Janeiro, RJ - Brazil}\author{Jos\'e Pimentel de Lima}
\affiliation{Departamento de F\'\i sica, Universidade Federal do Piau\'i, Teresina, PI - Brazil.}
\author{Thereza Paiva}
\affiliation{Instituto de F\'\i sica, Universidade Federal do Rio de Janeiro, Rio de Janeiro, RJ - Brazil}%
\author{Mohammed ElMassalami}
\affiliation{Instituto de F\'\i sica, Universidade Federal do Rio de Janeiro, Rio de Janeiro, RJ - Brazil}%
\author{Raimundo R. \surname{dos Santos}}
\affiliation{Instituto de F\'\i sica, Universidade Federal do Rio de Janeiro, Rio de Janeiro, RJ - Brazil}%


\begin{abstract}
With the purpose of investigating coexistence between magnetic order and superconductivity, we consider a model in which conduction electrons interact with each other, via an attractive Hubbard on-site coupling $U$, and with local moments on every site, via a Kondo-like coupling, $J$. 
The model is solved on a simple cubic lattice through a Hartree-Fock approximation, within a `semi-classical' framework which allows spiral magnetic modes to be stabilized.
For a  fixed electronic density, $n_c$, the small $J$ region of the ground state ($T=0$) phase diagram displays spiral antiferromagnetic (SAFM) states for small $U$. 
Upon increasing $U$, a state with coexistence between superconductivity (SC) and SAFM sets in; further increase in $U$ turns the spiral mode into a N\'eel antiferromagnet.
The large $J$ region is a (singlet) Kondo phase.
At finite temperatures, and in the region of coexistence, thermal fluctuations suppress the different ordered phases in succession: the SAFM phase at lower temperatures and SC at higher temperatures; also, reentrant behaviour is found to be induced by temperature.
Our results provide a qualitative description of the competition between local moment magnetism and superconductivity in the borocarbides family.

\end{abstract}

\date{\today}

\maketitle

\section{\label{sec1}Introduction}

Coexistence between magnetic order and superconductivity (SC) has been experimentally observed in many classes of materials, such as high-temperature cuprates \cite{Mukuda12,Scalapino12}, pnictides \cite{Stewart11,Scalapino12,Si16}, heavy fermions \cite{Pfleiderer09,Weng16,Steglich16}, and quaternary rare-earth compounds, such as the borocarbides \cite{Cava94,Canfield98, Muller2001,Muller2002,Gupta2006,Wolowiec15}.
Despite the many experimental advances in characterising these families of materials, a global description of the competition and interplay of both phenomena is still a matter of debate.
From the theoretical side, a unifying microscopic description of this coexistence is in its infancy. 
This may be attributed, to some extent, to the diversity both of the pairing mechanisms (electron-phonon or magnetic) and of the magnetic degrees of freedom (itinerant or localised). 
Nonetheless, one can gain considerable insight by investigating simplified models in which the basic physical processes are highlighted. 
For instance, the competition between superconductivity and magnetism in cuprates and iron pnictides has been extensively studied through the $t$-$J$ model 
\cite{Rodriguez11,Jedrak11,Abram13,Rodriguez14,Spalek17,Zegrodnik17}, 
while for heavy fermion compounds both the Periodic Anderson Model (PAM) 
\cite{Araujo01,Sacramento03,Sacramento10,Howczak13,Wu15,Wysokinski16} 
and the Kondo Lattice Model (KLM) 
\cite{Xavier08,Yu12,Bodensiek13,Asadzadeh14,Yu14,Lenz17} 
are more suitable.
When dealing with manifestly localised magnetism, such as in the quaternary borocarbides, one must consider localised moments on every site (or on a subset of sites, such as layers) of a lattice, interacting with each other via the conduction electrons; these, in turn, are subjected to a pairing interaction.
In our case, further simplifications are adopted, with the pairing tendency being provided by an on-site attractive (Hubbard) interaction, $U$ \cite{Micnas1990}, and 
the tendency towards magnetism being provided by the Kondo coupling, $J$, between the local moment and a conduction electron; 
one expects that effects arising from both the Ruderman-Kittel-Kasuya-Yosida (RKKY) interaction and the Kondo screening are contained in the Kondo coupling between conduction and localised electrons \cite{Hewson1993}.
This Kondo--attractive-Hubbard model (KAHM) was proposed in Ref.\,\cite{Bertussi09}, and studied in one dimension by means of density-matrix renormalisation group.
It was found that coexistence between spiral magnetic phases and superconductivity can indeed be found in the ground state for a range of parameters. 
For a fixed value of $U$, a sequence of magnetic modes is found as $J$ increases, ranging from (antiferromagnetic-like) spin-density waves up to a ferromagnetic state, though the latter was not found to coexist with superconductivity for any value of $U$; see Ref.\,\onlinecite{Bertussi09}.
A variant of this model, with a \emph{ferromagnetic} Kondo coupling, was investigated \cite{Karmakar16} in two dimensions, mostly in the ground state: a myriad of magnetic phases, coexisting or not with SC, was also observed.
Interestingly, sequences of magnetic modes have been observed in some quaternary borocarbides (QBC) with the general formula $RT_{2}$B$_{2}$C, in which $R$ is a rare earth element and $T$ is a transition metal \cite{Lynn1997,Muller2001,Muller2002,Gupta2006}.
In particular, alloying the transition metal, as in, e.g., Tb(Co$_x$Ni$_{1-x}$)$_2$B$_2$C or Ho(Co$_x$Ni$_{1-x}$)$_2$B$_2$C, yields sequences of magnetic modes, and may even lead to suppression of superconductivity, as $x$ is varied \cite{Schmidt1997,ElMassalami2012,ElMassalami2013,ElMassalami2014}. 

In view of these stimulating results, an investigation of this model in higher dimensions and at finite temperatures is definitely called for.
With this in mind, here we investigate the three-dimensional version of this model within a Hartree-Fock (HF) approximation.
Although this mean-field approach only takes into account local correlations, it captures general trends, especially in high dimensions. 
We note that we use a `semi-classical' approach \cite{Costa17a} to analyse the magnetic ordering, thus
allowing for spiral magnetic states to be stabilised; this approach has unveiled a multitude of magnetic phases in the Kondo lattice model \cite{Costa17a}.
We are therefore able to discuss the interplay between magnetic modes and superconductivity both in the ground state and at finite temperatures.

This paper is organized as follows: the model and the method are presented in Sec.\,\ref{sec2} (calculational details are left for the Appendix).
Sections \ref{sec3} and \ref{sec4} respectively discuss the results in the ground state and at finite temperatures. 
Finally, we discuss and summarize our findings in Sections \ref{sec:discussion} and \ref{sec:concl}, respectively.


\section{\label{sec2}Model and Method}

The Hamiltonian for the Kondo--attractive-Hubbard model \cite{Bertussi09} can be written as 
\begin{equation}
\mathcal{H} = \mathcal{H}_{K} + \mathcal{H}_{U},
\label{hamil}
\end{equation} 
with
\begin{equation}
\mathcal{H}_{K} = -t\!\sum_{\langle i, j \rangle, \sigma} \big( c^{\dagger}_{i\sigma} c^{\phantom{\dagger}}_{j\sigma} + \mathrm{H.c.} \big) + J \sum_{i} \mathbf{S}_{i}\! \cdot \mathbf{s}^{c}_{i}
\label{hamilKLM}
\end{equation} 
and
\begin{equation}
\mathcal{H}_{U} =  - U \sum_{i} n_{i \uparrow} n_{i \downarrow} ,
\label{hamilHub}
\end{equation}  
where the sums run over sites of a simple cubic lattice, with $\langle i,j \rangle$ denoting nearest-neighbour sites, $c^{\dagger}_{i \sigma}$ and $c^{\phantom{\dagger}}_{i \sigma}$ create and annihilate an electron on site $i$ with spin $\sigma$, H.c.\,denotes the hermitian conjugate of the previous expression, and $\mathbf{S}_{i}$ and $\mathbf{s}^{c}_{i}$ are the spin operators for the local moments and conduction electrons, respectively.  
Therefore, $\mathcal{H}_{K}$ is the Kondo lattice Hamiltonian describing the interplay between delocalisation and magnetic ordering (we take the exchange as $J > 0$).
The term $\mathcal{H}_{U}$ corresponds to a local attractive interaction between conduction electrons, with coupling strength $-U < 0$; $ n_{i \sigma} = c^{\dagger}_{i \sigma} c^{\phantom{\dagger}}_{i \sigma} $ is the number operator of conduction electrons on site \textit{i} and spin $\sigma$.

It is worth mentioning that the standard Bardeen-Cooper-Schrieffer (BCS) pairing mechanism differs from the contents of the attractive-Hubbard term, Eq.\,\eqref{hamilHub}.
While in the former only electrons near the Fermi surface are paired, in the latter all electrons within the Fermi sphere feel the attractive interaction.
Nonetheless, the attractive-Hubbard model shares some similarities with the BCS model, in particular for weak coupling (see, e.g., Ref.\,\cite{Micnas1990}), which is the regime we consider throughout this work. 

Within a Hartree-Fock approximation, we decouple the quartic terms, leading to a quadratic Hamiltonian, with effective fields to be determined self-consistently.
To this end, we write the spin operators in a fermionic basis as 
\begin{equation} \label{Sf}
\mathbf{S}_{i} = \frac{1}{2}\sum_{\alpha, \beta = \pm } f^{\dagger}_{i \alpha} \boldsymbol{\sigma}_{\alpha, \beta}f^{\phantom{\dagger}}_{i \beta} ,
\end{equation}
and
\begin{align} \label{Sc}
\mathbf{s}^{c}_{i} = \frac{1}{2}\sum_{\alpha, \beta = \pm} c^{\dagger}_{i \alpha} \boldsymbol{\sigma}_{\alpha, \beta} c^{\phantom{\dagger}}_{i \beta} ,
\end{align}
with $ \boldsymbol{\sigma}_{\alpha, \beta}$ denoting the elements of the Pauli matrices, and $f^{\dagger}_{i\sigma}$ ($f^{\phantom{\dagger}}_{i\sigma}$) being creation (annihilation) operators for localised electrons.

Following the procedure outlined in the Appendix \ref{Ap}, the Hamiltonian of Eq.\,\eqref{hamil}, 
becomes
\begin{widetext}
\begin{align} 
\nonumber
\mathcal{H}_{MF} = & \sum_{\mathbf{k} \sigma} (\epsilon_{\mathbf{k}} - \tilde{\mu})\, c^{\dagger}_{\mathbf{k} \sigma} c^{\phantom{\dagger}}_{\mathbf{k} \sigma}
+ \bigg( \frac{J m_{f}}{2} - U m_{c} \bigg) \sum_{\mathbf{k}} \big( c^{\dagger}_{\mathbf{k} \uparrow} c^{\phantom{\dagger}}_{\mathbf{k} + \mathbf{Q} \downarrow} + \mathrm{H.c.} \big)
- \frac{J m_{c}}{2} \sum_{\mathbf{k}} \big( f^{\dagger}_{\mathbf{k} \uparrow} f^{\phantom{\dagger}}_{\mathbf{k} + \mathbf{Q} \downarrow} + \mathrm{H.c.} \big)
%
\\ \nonumber &
+ \frac{3}{4}JV  \sum_{\mathbf{k} \sigma} \big( c^{\dagger}_{\mathbf{k} \sigma} f^{\phantom{\dagger}}_{\mathbf{k} \sigma} + \mathrm{H.c.} \big)
+ \frac{J V'}{4}  \sum_{\mathbf{k}} \big( c^{\dagger}_{\mathbf{k} \uparrow} f^{\phantom{\dagger}}_{\mathbf{k} + \mathbf{Q} \downarrow} + c^{\dagger}_{\mathbf{k} + \mathbf{Q} \downarrow} f^{\phantom{\dagger}}_{\mathbf{k} \uparrow} + \mathrm{H.c.} \big)
+ \epsilon_{f} \sum_{\mathbf{k} \sigma} f^{\dagger}_{\mathbf{k} \sigma} f^{\phantom{\dagger}}_{\mathbf{k} \sigma} 
%
\\ &
 - t\Delta \sum_{\mathbf{k}} \big( c^{\dagger}_{\mathbf{k} \uparrow} c^{\dagger}_{-\mathbf{k} \downarrow} +  \mathrm{H.c.} \big)
 + N \bigg[ \mu n_{c} - \epsilon_{f} + J m_{c} m_{f} + \frac{3}{2} J V^{2} - \frac{1}{2} J V'^{2} + \frac{U n_{c}^{2}}{4} + 
 \frac{(t\Delta)^2}{U} - U m_{c}^{2} - \frac{U n_{c}}{2} \bigg],
\label{hamilKaH}
\end{align}
\end{widetext}
where $\tilde{\mu}$ and $\epsilon_{\mathbf{k}} = -2t\big[ \cos(k_{x}) + \cos(k_{y}) + \cos(k_{z}) \big]$ are the chemical potential and the dispersion of the bare conduction electrons, respectively, while $n_{c}$ is their density; the lattice spacing is taken as unity. 
The localised electrons are dispersionless, with $\epsilon_{f}$ being their renormalized contribution to the energy. 
In our approach, the hybridization between $c$ and $f$ bands is represented by the effective fields $V$ and $V'$, 
which are defined through 
\begin{equation}\label{singlet_hyb1}
	V\equiv \langle V_{ic}^0\rangle = \langle {V^{0}_{i f}}^{\dagger}\rangle 
		= \frac{1}{2}\sum_{\alpha, \beta = \pm} \langle c^{\dagger}_{i \alpha} \mathbb{1}_{\alpha, \beta} f^{\phantom{\dagger}}_{i \beta}\rangle,
\end{equation}
referred to as the singlet hybridization, and through the triplet hybridization,
\begin{equation}\label{triplet_hyb_mean_v1}
\langle \mathbf{V}_{i c} \rangle  = \langle  {\mathbf{V}_{i f}}^{\dagger} \rangle = V' \big[ \cos \left(\mathbf{Q}\!\cdot\! \mathbf{R}_{i}\right), \sin \left(\mathbf{Q}\!\cdot\! \mathbf{R}_{i}\right),0 \big],
\end{equation}
with
\begin{eqnarray}\label{triplet_hyb1}
\mathbf{V}_{i c} = \mathbf{V}^{\dagger}_{i f} = \frac{1}{2} \sum_{\alpha, \beta = \pm} c^{\dagger}_{i \alpha} \boldsymbol{\sigma}_{\alpha, \beta} f^{\phantom{\dagger}}_{i \beta},
\end{eqnarray}
and 
\begin{equation} 
	\mathbf{Q}=(q_{x}, q_{y}, q_{z}).
\label{Q1}
\end{equation}

The magnetic ordering is probed by the wave vector $\mathbf{Q}$ and the amplitudes $m_f$ and $m_c$
(i.e. the magnetisations associated with the sublattice comprised of local moments and conduction electrons, respectively),
which describe the average magnetisations,
\begin{equation}
	\label{Si1}
	\langle \mathbf{S}_{i} \rangle = m_{f} \big[ \cos \left(\mathbf{Q}\!\cdot\! \mathbf{R}_{i}\right), 
		\sin \left(\mathbf{Q}\!\cdot\! \mathbf{R}_{i}\right), 0 \big]
\end{equation}
and
\begin{equation}
	\label{Sci1}
	\langle \mathbf{s}^{c}_{i} \rangle = 
	-m_{c} \big[ \cos \left(\mathbf{Q}\!\cdot\! \mathbf{R}_{i}\right), \sin \left(\mathbf{Q}\!\cdot\! \mathbf{R}_{i}\right),0 \big],
\end{equation}
where $\mathbf{R}_{i}$ is the vector position of site $i$ on the lattice.
At this point we note that while the lattice we consider here is \emph{structurally} three-dimensional (cubic), the above \emph{ansatze} for the magnetic degrees of freedom introduce an anisotropy in the \emph{magnetic lattice.}

And, finally, $\Delta$ is the (dimensionless) $s$-wave superconducting order parameter, 
\begin{equation}
	\Delta=\frac{U}{t} \langle c^{\dagger}_{i \uparrow} c^{\dagger}_{i \downarrow} \rangle 
		= \frac{U}{t} \langle c_{i \downarrow} c_{i \uparrow} \rangle,
\end{equation}
where the site-dependence in $\Delta$ was omitted due to the homogeneity assumption within the Hartree-Fock approximation.
We note that the local attractive interaction favours on-site pairing 
\cite{Micnas1990,Paiva2004}
i.e. s-wave symmetry.
By the same token, one expects the Kondo-like term to contribute to inter-orbital pairing (e.g. $c^{\dagger}f^{\dagger}$ \cite{Spalek88}), which plays an important role in unconventional superconductivity \cite{Cyrot86,Gehring94} in the context of heavy fermion materials, such as CeCoIn$_{5}$ 
\cite{Spalek88,Howczak13,Masuda13,Masuda15,Wysokinski16}.
However, here we are dealing with well localised magnetism, such as Tm, Er and Ho-based materials, so that the contribution from unconventional channels may be disregarded in a first approach. 

The effective fields provided by the mean-field approximation are $\tilde{\mu}$, $\epsilon_{f}$, $V$, $V'$, $m_{f}$, $m_{c}$, $\mathbf{Q}$ and $\Delta$. Actually, $\tilde{\mu}$ and $\epsilon_{f}$ are effective fields included as Lagrange multipliers, in order to fix (in average) the electronic density and the number of local moments per site. The mean-field Hamiltonian can now be diagonalised using Nambu spinor representations, with eight-component spinors, leading to a two-fold degenerated bands $E^{n}_{\mathbf{k}}$, ($n = 1,\ldots, 8$). Then, the Helmholtz free energy is
\begin{equation}
\label{Helmholtz}
F = -\frac{1}{\beta} \sum_{n, \mathbf{k}} \ln \big( 1 + e^{-\beta E^{n}_{\mathbf{k}}} \big) + const ,
\end{equation}
where $\beta=1/k_{B}T$. 

Minimizing the Helmholtz free energy
\begin{align} \label{selfcon}
\nonumber  &\bigg\langle \frac{\partial F}{\partial \tilde{\mu}} \bigg\rangle = \bigg\langle \frac{\partial F}{\partial \epsilon_{f}} \bigg\rangle = \bigg\langle \frac{\partial F}{\partial V} \bigg\rangle = \bigg\langle \frac{\partial F}{\partial V'} \bigg\rangle =  \\ 
 & \bigg\langle \frac{\partial F}{\partial m_{f}} \bigg\rangle = \bigg\langle \frac{\partial F}{\partial m_{c}} \bigg\rangle = \bigg\langle \frac{\partial F}{\partial q_{\alpha}} \bigg\rangle = \bigg\langle \frac{\partial F}{\partial \Delta} \bigg\rangle = 0 ,
\end{align}
we are able to determined self-consistently the effective fields. The resulting nonlinear coupled equations are solved numerically using standard library routine packages, with aid of the Hellmann-Feynman theorem. 
One should have in mind that it is the combination of the hybridisation terms, \eqref{singlet_hyb1} and \eqref{triplet_hyb1}, of the semi-classical \emph{ansatze}, \eqref{Si1} and \eqref{Sci1}, and with the minimisation also with respect to $\mathbf{Q}$, that allows the description  of spiral phases, as verified in the context of the Kondo lattice model.\cite{Costa17a} 


\section{\label{sec3} Ground State properties}

When $U=0$, the Hamiltonian reduces to the KLM, which is known to  be an insulator at half filling for all $J$, with a N\'eel ground state for $J/t<(J/t)_c\approx 4$, and a spin singlet for $J/t>(J/t)_c$~\cite{Lacroix1979}.
Away from half filling, the system is metallic and spiral states may be stabilised, as recently reported for the two-dimensional case \cite{Costa17a}; we will see below that indications of similar sequences of spiral phases are also found in the present three-dimensional case, though we have not pursued here an analysis as detailed as that of Ref.\,\onlinecite{Costa17a}.
The strong-coupling limit, $J/t\gg1$, corresponds to the Kondo phase, with  conduction electrons strongly hybridised with local moments, thus forming singlets.
In the opposite limit, $J=0$, Eq.\,\eqref{hamil} reduces to the attractive Hubbard model whose ground state is superconducting for any $U>0$ and any band filling \cite{Micnas1990,dosSantos94b}.
When both $J$ and $U$ are non-zero, these two tendencies compete with each other, and numerical solutions to the nonlinear coupled equations, Eqs.\,\eqref{selfcon}, are sought for a given set of control parameters, ($n_{c}$, $J$, $U$), with $T=0$.

\begin{figure}[t]
\includegraphics[scale=0.3]{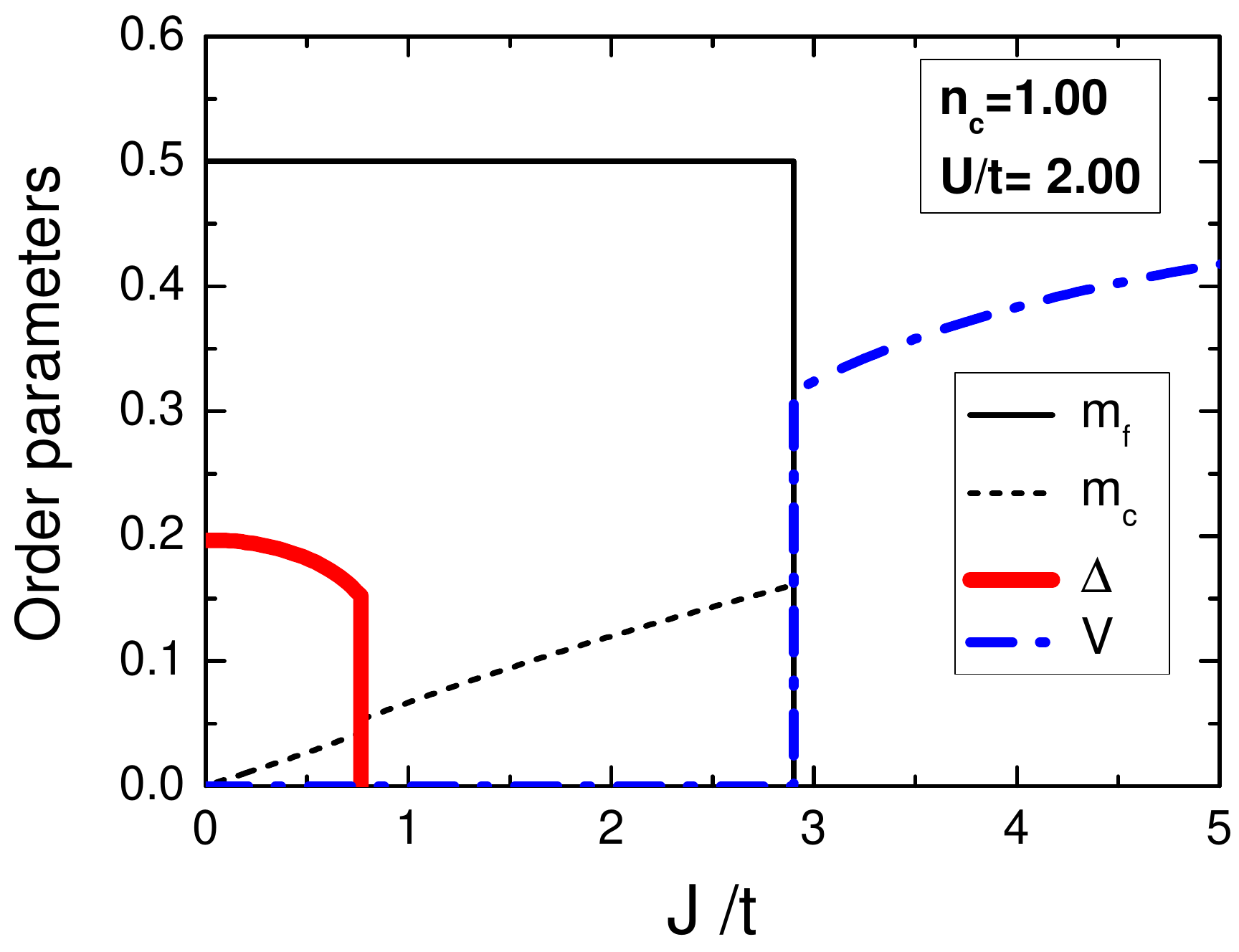}
\caption{(Colour online) 
Local moment
sublattice
magnetisation, $m_f$, conduction electron
sublattice
magnetisation, $m_c$, superconducting gap, $\Delta$, and the hybridisation amplitude, $V$, (see text for definitions) 
as functions of $J/t$, for fixed $U/t=2$ and at half filling. The magnetic wave-vector $\mathbf{Q}=(\pi, \pi, \pi)$ minimizes the free energy for all values of $J/t$.}
\label{order_100} 
\end{figure}

Let us first discuss the case of a half-filled conduction band, $n_c=1$,  in which both the antiferromagnetic (AFM) and Kondo phases are insulating; in what follows, all borders around the SC+AFM phase correspond to superconductor-insulator phase transitions.
Figure \ref{order_100} shows the different order parameters as functions of $J/t$, for $U/t=2$ and at half filling. 
The superconducting order parameter, $\Delta$, decreases with $J/t$ and vanishes abruptly at $(J/t)_{c1}\approx 0.8$.
The magnetisation of the localised spins corresponds to $\mathbf{Q}=(\pi,\pi,\pi)$, and its magnitude, $m_f$, is not affected by the exchange coupling as long as $J/t<(J/t)_{c2}\approx 2.9$, when it drops abruptly to zero. 
Therefore N\'eel antiferromagnetism (of the localised spins) coexists with superconductivity up to $(J/t)_{c1}$. 
Further, the conduction electron magnetisation amplitude, $m_c$, increases with $J/t$, suffers a tiny burst when superconductivity disappears, and carries on increasing up to $(J/t)_{c2}$, where, similarly to $m_f$, drops to zero. 
The picture that emerges is that an increase in $J/t$ leads to a pair-breaking effect followed by an antiferromagnetic coupling of the released electron to the local magnetic moment; for $(J/t)_{c1}<J/t<(J/t)_{c2}$ the pairing tendency disappears, and the Kondo lattice regime of polarised electrons coexisting with the N\'eel background of local moments remains. 
Only above $(J/t)_{c2}$ the conduction electrons become hybridised with the local moments (the Kondo phase), thus suppressing magnetic order; at $(J/t)_{c2}$ this is signalled by both the vanishing of $m_c$ and $m_f$, and the jump of $V$ from zero to a finite value.   
For completeness, from our experience with the KLM\cite{Costa17a} we note that the triplet hybridisation only plays some role when there is coexistence between magnetism and the Kondo phase. 
We have checked that the range of parameters for which this occurs is in fact very narrow in three dimensions, and shrinks even further as $|U|$ increases; for instance, for $n_c=0.9$ and $3 < J/t < 4$ the AFM+Kondo phase disappears already 
for $U/t\approx 0.44$.  

\begin{figure}[t]
\includegraphics[scale=0.3]{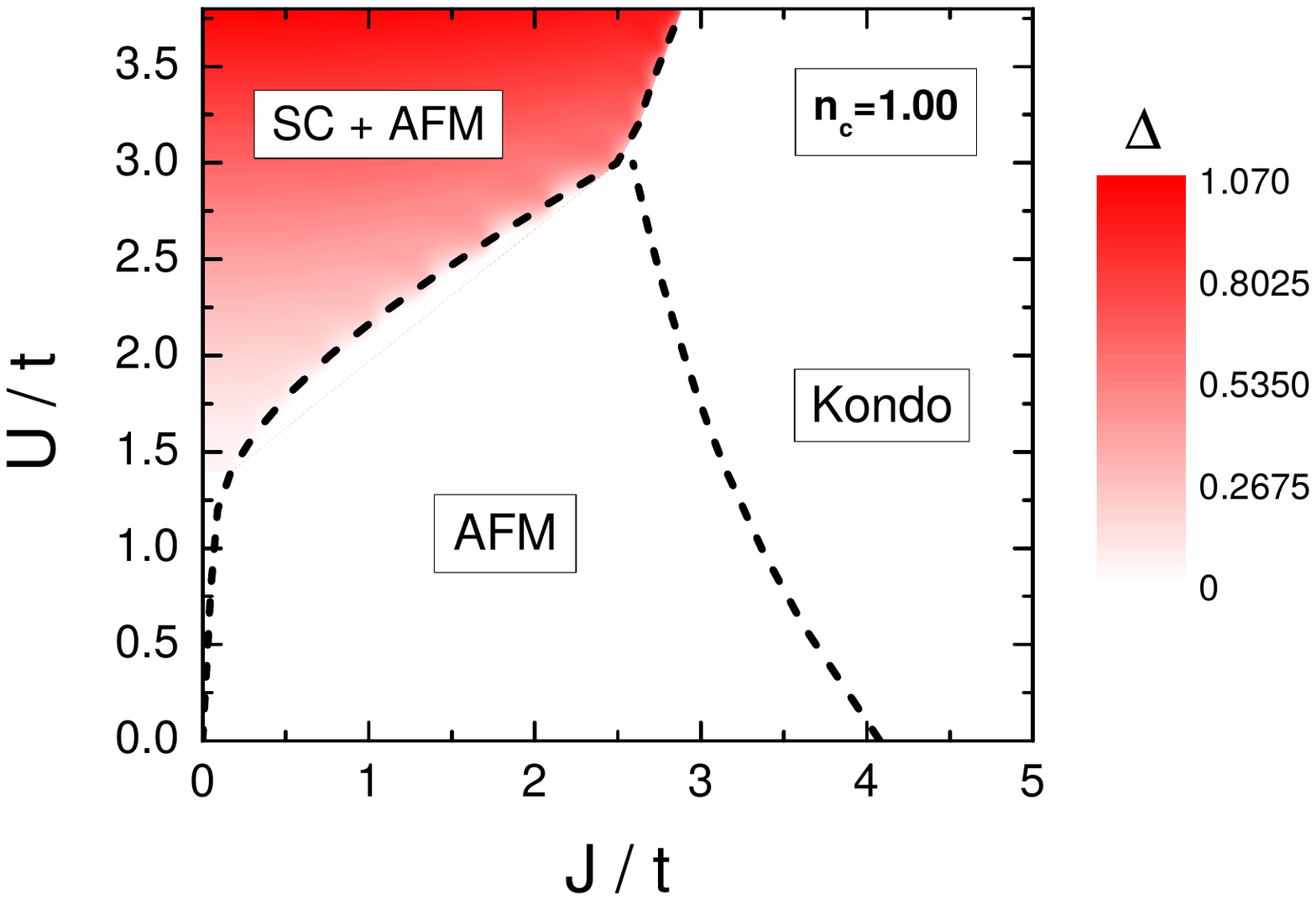}
\caption{(Colour online) Ground state phase diagram at half filling, showing the antiferromagnetic (AFM), Kondo, and coexisting superconducting and AFM phases (SC$+$AFM).
The dashed lines represent first order phase transitions, while the magnitude of the superconducting gap, $\Delta$, is mapped by the intensity of the shading in the SC$+$AFM region, with the numerical scale shown to the right of the plotting area. 
Throughout this paper we adopt the following convention when displaying phase diagrams: continuous and dashed lines respectively denote continuous and first order transitions.}
\label{diag100} 
\end{figure}

By performing the same analysis for other values of $U$, we obtain the phase diagram shown in Fig.\,\ref{diag100}. 
Note that $(J/t)_{c1}$ increases with $U$, while $(J/t)_{c2}$ decreases with $U$; that is, the purely insulating AFM region shrinks with increasing $U$, concomitant with an increase in both the coexisting region and the Kondo phase, so that beyond $U/t\approx 3$ the system cannot sustain pure antiferromagnetic order: it is either an antiferromagnetic superconductor or a Kondo insulator, depending on $J/t$.  
Figure \ref{diag100} also indicates that at half filling $U$ cannot change the magnetic mode into a spiral state.  

\begin{figure}[t]
\includegraphics[scale=0.30]{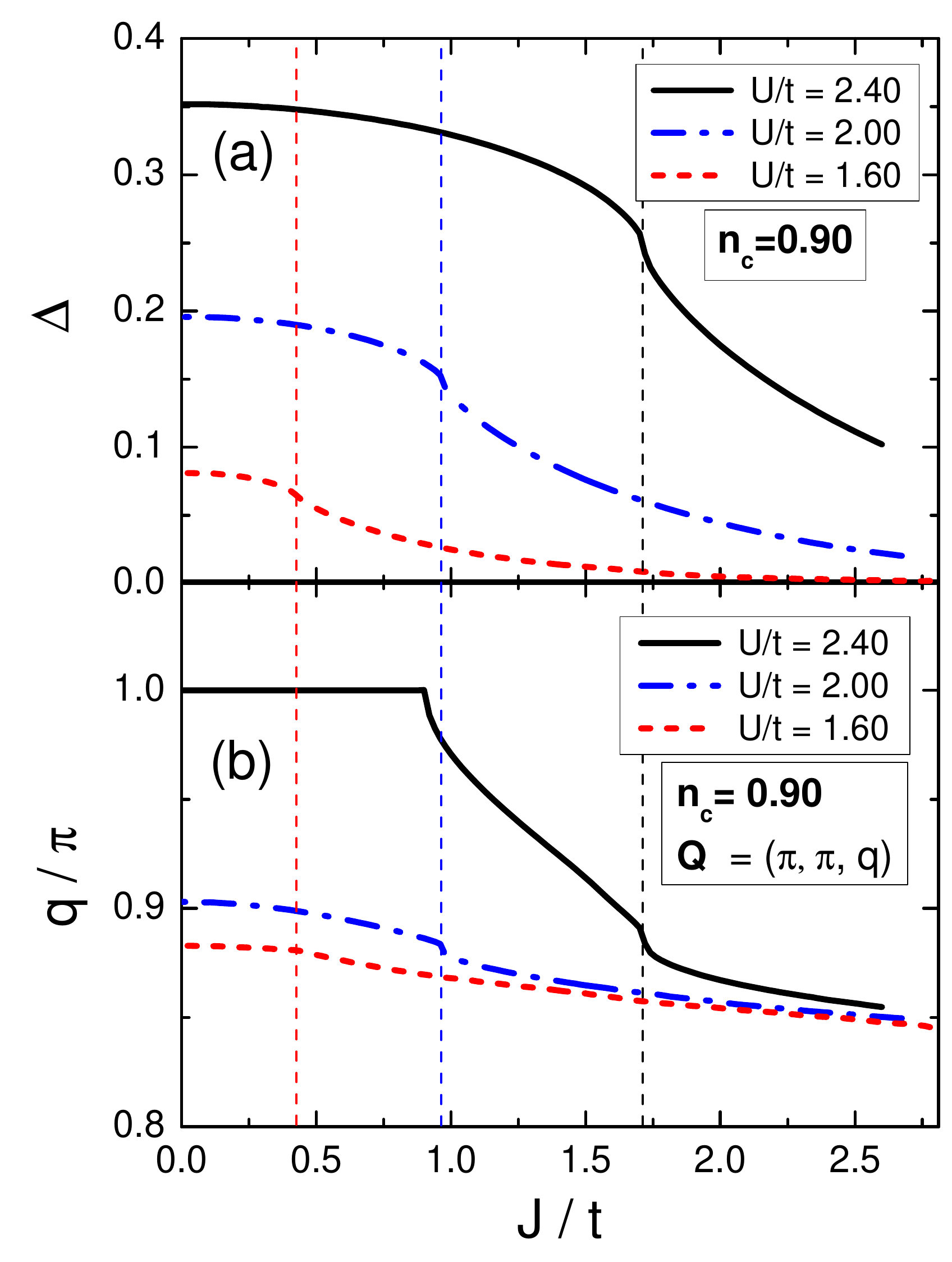}
\caption{(Colour online) Superconducting gap (a) and $z$-component of the magnetic wave vector (b) as functions of the Kondo exchange coupling $J/t$, for three different values of the local attraction $U/t$. The vertical dashed lines locate the inflection points $(J/t)^\times$ for each value of $U/t$ in panel (a), and are prolonged into panel (b) to correlate with the inflections in $q$; see text.}
\label{Deltas} 
\end{figure}

Away from half filling the system behaves quite differently, as illustrated in Fig.\,\ref{Deltas}(a) for the small doping regime, $n_{c}=0.90$, and for three different values of $U/t$.
While increasing $J/t$ is still detrimental to superconductivity, as signalled by the decrease in $\Delta$, here the superconducting gap does not drop to zero abruptly: beyond some inflection point, $J^\times(U)$, the gap decays to zero with an exponential tail; from now on, we will refer to this regime as  
`weakly superconducting' (WSC) \footnote{\label{foot25}  The meaning of \emph{weakly superconducting} in the context of this paper has no connection with the more common usage of the term, denoting the type of superconductivity occurring on a surface or a grain boundary; here it simply denotes a weakening of the gap.} but, as Fig.\,\ref{Deltas}(a) shows, this change in behaviour is to be regarded as a crossover, not as a phase transition.
This exponential behaviour can be attributed to the fact that the superconducting gap for the attractive Hubbard model is expected to behave as \cite{Micnas1990} 
\begin{equation}
	\widetilde{\Delta} \simeq W \exp \left(-\frac{1}{N_0U}\right),
\label{eq:Delta}
\end{equation}
where $\widetilde{\Delta}$ denotes the gap in energy units, $W$ is the band width (recall that in standard BCS theory the scale for the gap is set by the Debye energy, $\hbar\omega_D$), and $N_0$ is the density of states at the Fermi energy in the normal phase.
Away from half filling the normal state is metallic, so $N_0$ is finite. 
Further, an examination of the band structure for the KLM (not shown) reveals that $N_0$ decreases with increasing $J/t$, a trend which should also hold for a fixed nonzero $U$; the gap therefore decreases exponentially with $J/t$.  
This argument also explains why the gap vanishes abruptly at half filling: the normal phase is insulating, so that $N_0=0$ by decreasing $J/t$  from either the Kondo- or the AFM phases. 
Figure \ref{Deltas}(b) shows the behaviour of the magnetic wave vector [actually, of the $z$-component in $\mathbf{Q}=(\pi,\pi,q)$] with the Kondo coupling: for a fixed $U/t$, $q$ decreases with $J/t$, and displays an inflection at the same $(J/t)^\times$ as $\Delta$ does.
It should also be noted that increasing $J/t$ causes $q$ to approach the same finite value, independent of $U/t$, just before the system enters the Kondo phase. 
This seems to indicate that before forming singlets, the local moments lock into a magnetic mode which (in the ground state) depends solely on the density of conduction electrons. 
 
\begin{figure}[t]
\includegraphics[scale=0.3]{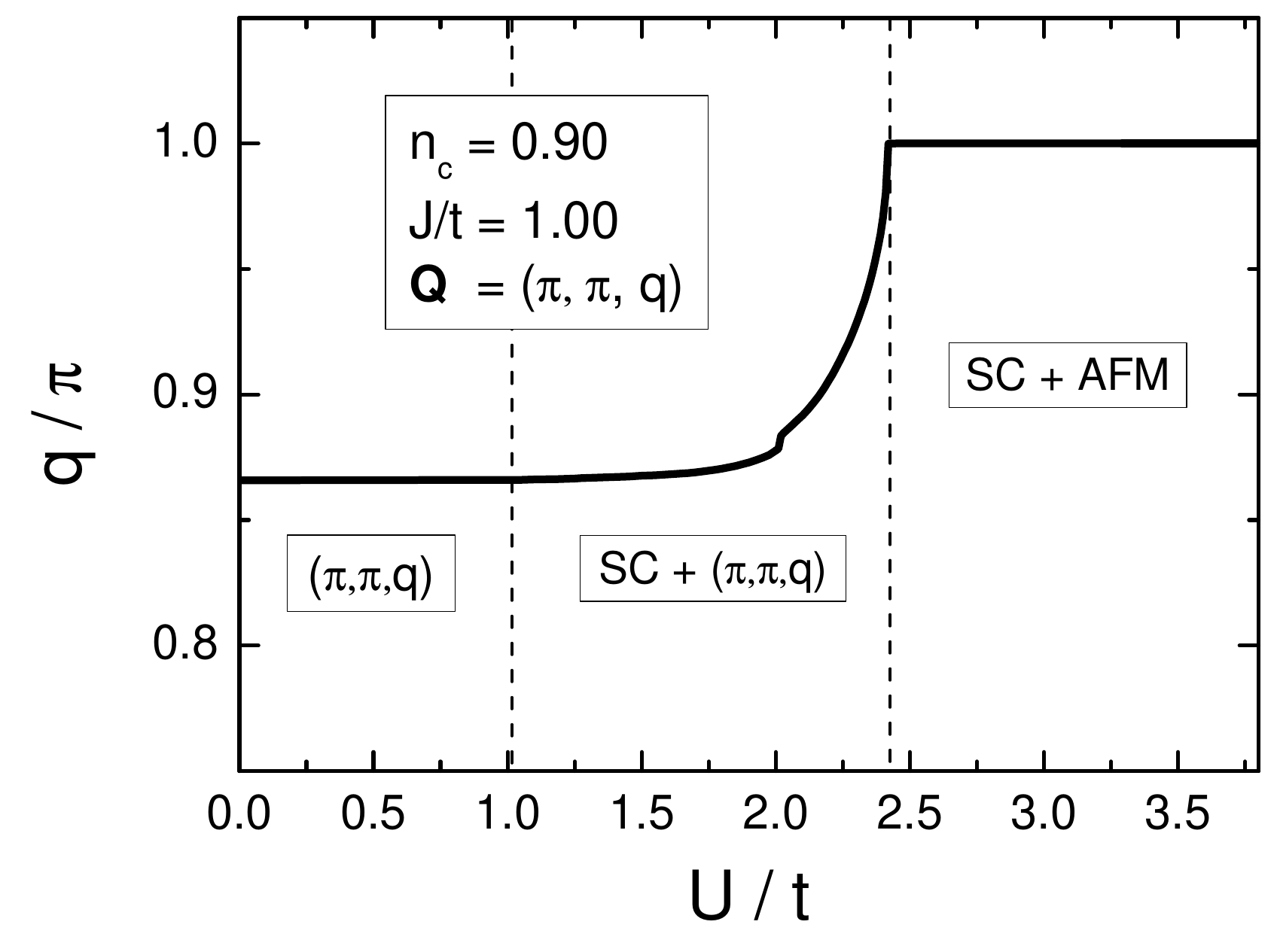}
\caption{The \textit{z}-component of the magnetic wave vector as a function of $U/t$, for a fixed $J/t$, and away from half filling. The dashed vertical lines indicate change in regimes, for the choice of parameters shown: superconductivity only sets in for 
$U/t\geq 1$, and the AFM mode is only stable for $U/t\gtrsim 2.4$}
\label{qU} 
\end{figure}

From Fig.\,\ref{Deltas}(b) we also see that increasing $U/t$ with fixed $J/t$ has the opposite effect, namely to increase $q$. 
Figure \ref{qU} provides a closer look at the evolution of $q$ across the different regimes.
We note that when the system is either in the normal or WSC regimes, the $U$ dependence of the magnetic wave vector is quite feeble; by contrast, when the system is in the SC regime, the increasing pairing interaction causes a rapid increase in $q$ towards $q=\pi$, thus leading to N\'eel magnetism coexisting with superconductivity. 
This can be understood as due to a stronger binding of the local pairs, which in turn polarizes the local moments into a N\'eel state: any spiral magnetic mode would cost more energy to delocalise the bound pairs; the effective outcome is a smaller pair-breaking effect on the conduction electrons.  

\begin{figure}[t]
\includegraphics[scale=0.32]{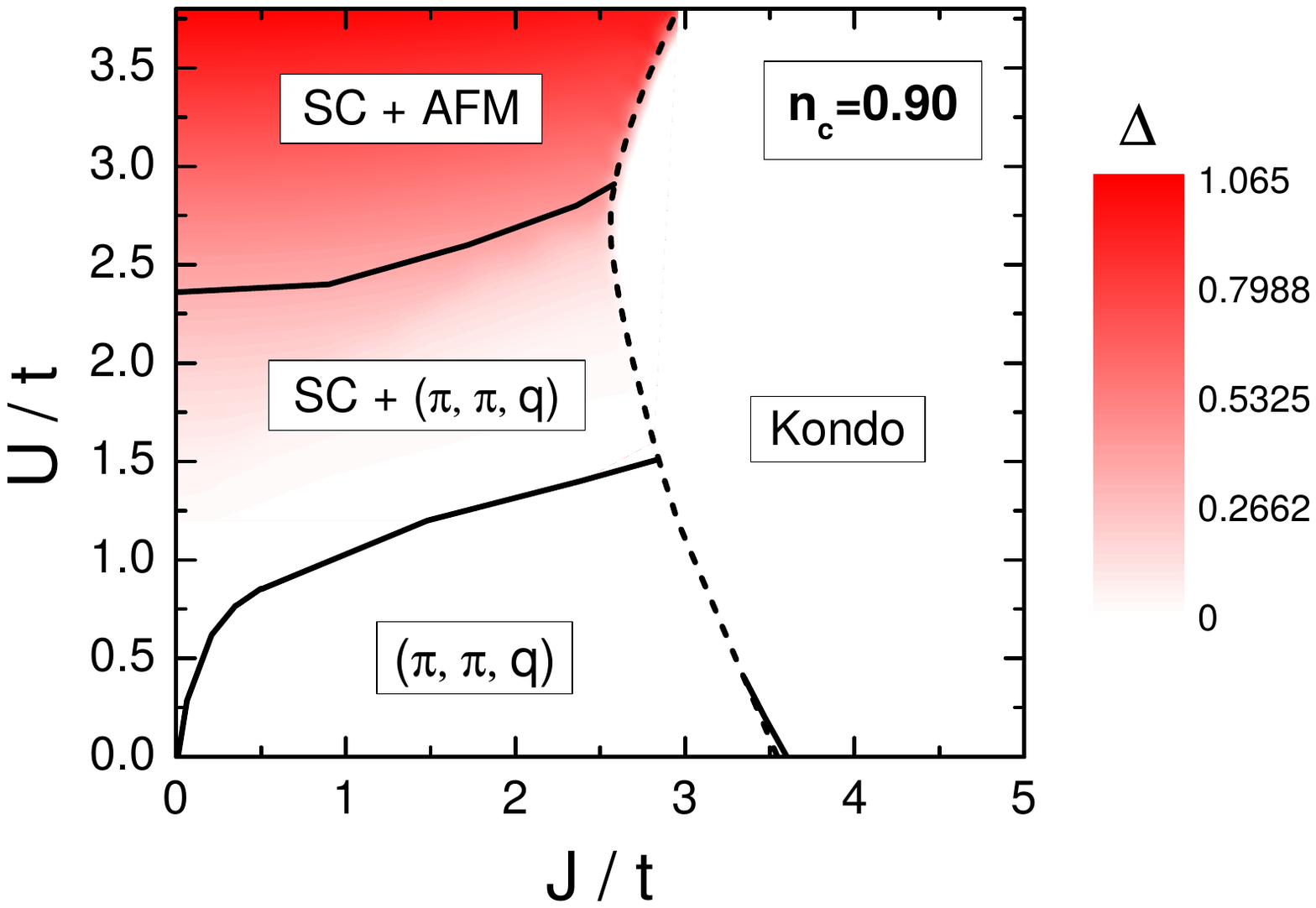}
\caption{(Colour online) Ground state phase diagram for low doping, $n_{c}=0.90$.\cite{Note1}
The magnitude of the superconducting gap, $\Delta$, is mapped by the intensity of the shading in the SC area, with the numerical scale shown to the right of the plotting area. The dependence of $q$ with $U/t$ and $J/t$ is illustrated in Figs.\,\ref{Deltas} and \ref{qU}.}
\label{diag090} 
\end{figure}

The phase diagram in Fig.\,\ref{diag090} summarises our results for $n_c=0.9$.\footnote{\label{foot21}  
We recall that the phase boundaries are determined by crossings in the lowest ground state energy, and they coincide with those obtained by the vanishing of the different order parameters; when approaching the spiral magnetic phase at small $U/t$, we observe $\Delta$ being smaller than a given tolerance, in accordance with the decreasing exponential behaviour.}
For small attractive coupling, spiral magnetic order appears without coexistence with SC. 
However, for larger values of $U/t$, the ground state displays coexistence between SC and magnetism, with the magnetic wave vector evolving from a spiral mode to AFM, when $U/t$ increases. 
On the other hand, for large exchange coupling, the Kondo phase is more stable than both SC and magnetism. 
The trends predicted here are in good agreement with those obtained through DMRG for the same model in one dimension:\cite{Bertussi09} 
one of the main features, namely the evolution through incommensurate magnetic modes induced by SC, is also present in our three-dimensional mean-field analysis.

\begin{figure}[t]
\includegraphics[scale=0.32]{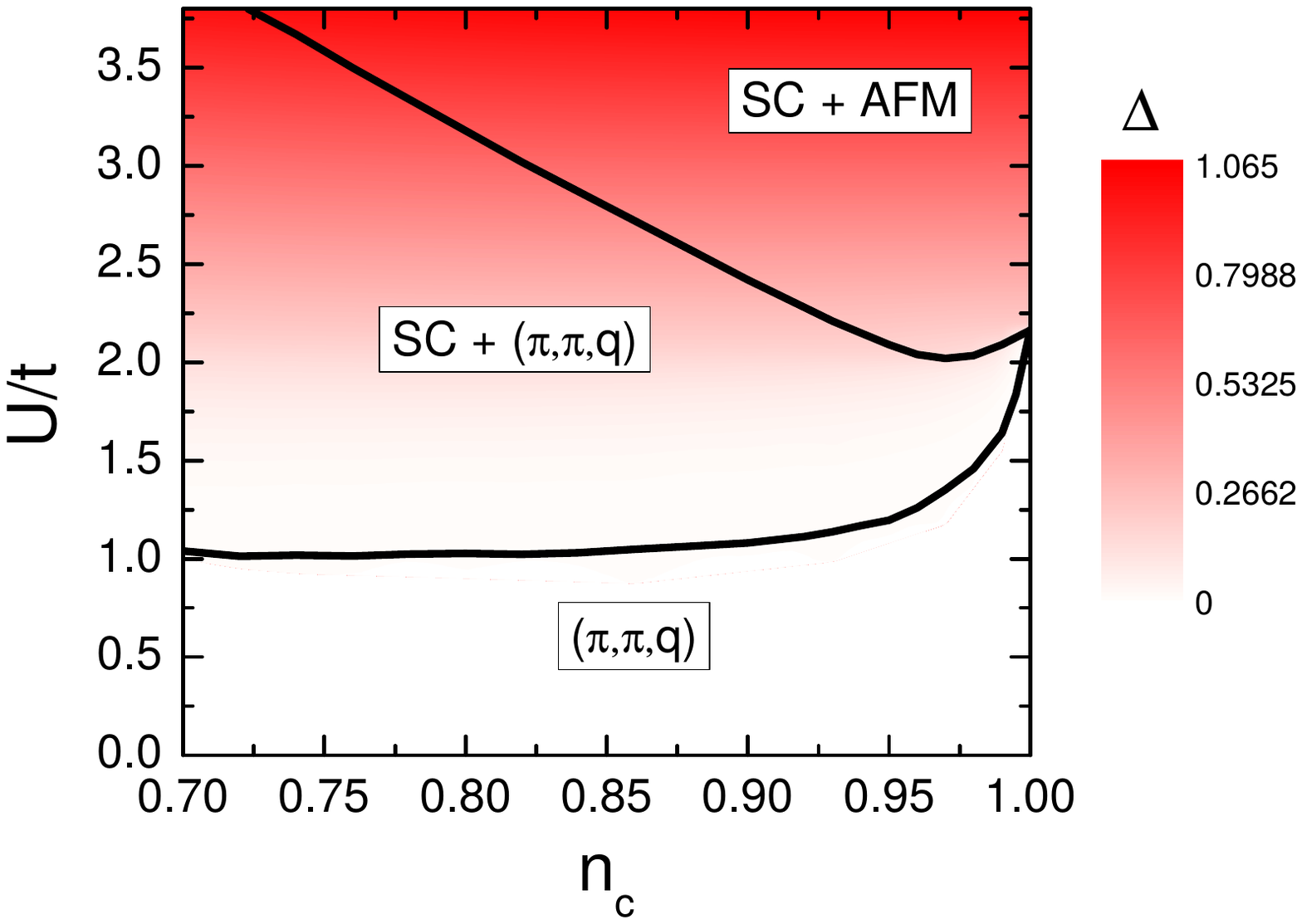}
\caption{(Colour online) Ground state phase diagram away from half filling, for fixed $J/t= 1.0$.
The magnitude of the superconducting gap, $\Delta$, is mapped by the intensity of the shading in the SC$+$AFM region, with the numerical scale shown to the right of the plotting area.}
\label{diag_nc} 
\end{figure}

The above analyses are repeated for other electronic densities, and we obtain the phase diagram $U/t\times n_c$, for fixed $J/t$, displayed in Fig.\,\ref{diag_nc}, for $n_{c} \geq 0.70$. 
We note that for small $U/t$, the ground state exhibits a spiral magnetic phase, with $\mathbf{Q}=(\pi, \pi, q)$, which is reminiscent of the KLM, since the attractive potential does not change the magnetic wave vector in the absence of SC, as showed in Fig.\,\ref{qU}.
When $U/t$ increases, SC sets in and coexistence with magnetism appears.
For larger values of $U/t$ the magnetic mode tends to AFM, as discussed before.
However, the value of $U$ at which the spiral mode changes to AFM depends quite strongly on the electronic density; the larger the doping, the larger is the value of $U/t$ required to enter the phase SC$+$AFM.
This results from two opposing tendencies. 
On the one hand, as one dopes the KLM away from half filling, a magnetic mode with $\mathbf{Q}=(\pi, \pi, q)$ changes continuously to $\mathbf{Q}=(\pi, \pi, 0)$;\cite{Costa17a} 
on the other hand, we have seen that increasing $U$ drives the magnetic mode towards $q=\pi$.
Therefore, the smaller the electronic density, the larger $U$ must be to induce the $\mathbf{Q}=(\pi, \pi, \pi)$ mode.  

It is also instructive to discuss the behaviour of the SC order parameter as a function of $n_{c}$ for different values of $J/t$, and fixed $U/t$, as shown in Fig.\,\ref{ncs} for $U/t=2$;
for comparison, the behaviour of $\Delta$ for the simple attractive Hubbard model ($J/t=0$) is also shown in Fig.\,\ref{ncs}.
While for $n_c\lesssim 0.9$, the gaps are merely shifted from the AHM case, we note that near half filling the behaviour changes significantly with increasing $J/t$: the SC order parameter is strongly suppressed as half filling is approached. 
This change in the behaviour of $\Delta$ occurs abruptly at $J/t \approx 0.75$, which is the transition point where SC is suppressed at half filling, for $U/t=2$; see Figs.\,\ref{order_100} and \ref{diag100}.
As we will see in the next Section, this reduction in $\Delta$ near half filling leads to a re-entrant-like behaviour in SC at finite temperatures.

At this point, we should comment on the order of the phase transitions.
Since mean-field approximations only take into account local fluctuations, they may be inadequate in the study of criticality and quantum phase transitions: some of the transitions which we have obtained as first-order, may become second-order if quantum fluctuations were incorporated in a more fundamental way.
Nonetheless, some first-order transitions are quite robust: very accurate mean-field approximations \cite{Peters15,Peters17}, a variational Gutzwiller approach \cite{Lanata08}, and  Monte Carlo methods \cite{Watanabe07,Asadzadeh13} all agree in predicting a first-order transition from AFM to Kondo phase for the KLM Hamiltonian \big[Eq.\,\eqref{hamilKLM}\big], usually associated with an abrupt change in the volume of the Fermi surface.


\section{\label{sec4} Finite Temperatures}

\begin{figure}[t]
\includegraphics[scale=0.3]{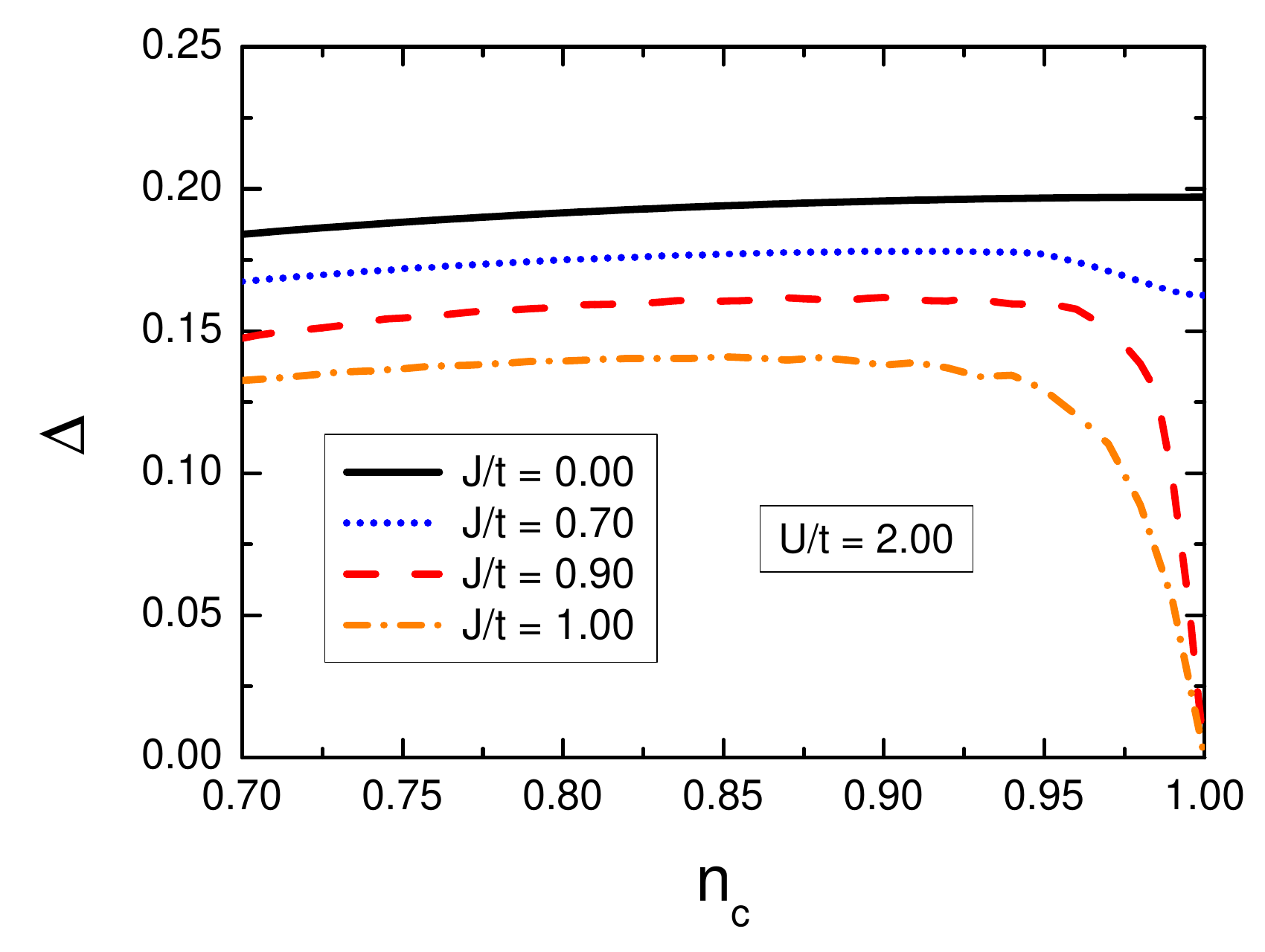}
\caption{(Color online) SC order parameter, for fixed $U/t=2$, as function of the electronic density. }
\label{ncs} 
\end{figure}

We now turn to discuss how the temperature affects the phases observed in the ground state. 
With the parameter space comprising of temperature, $T/t$, the couplings $J/t$ and $U/t$, and the band filling, $n_c$, we should impose some restrictions on the ranges of $J/t$ and $U/t$ involved, based on some experimental constraints. 
For instance, for the borocarbides the N\'eel temperatures,\footnote{
Here we use the term N\'eel temperature, $T_N$, in a broader sense, encompassing the critical temperature for spiral states as well.} $T_N$, are of the same order of magnitude as those of the critical temperatures for SC, $T_c$. 
We therefore consider ranges of $U/t$ and $J/t$ leading to values for $T_{N}$ and $T_{c}$ of similar magnitude; these ranges are narrowed down even further by choosing parameters such that superconductivity coexists with magnetism in the ground state, which, as we have seen, necessarily leads to spiral magnetic modes.
As we will see, this suffices to highlight several features arising from the competition between SC and magnetism which become more evident at finite temperatures.

\begin{figure}[t]
\includegraphics[scale=0.3]{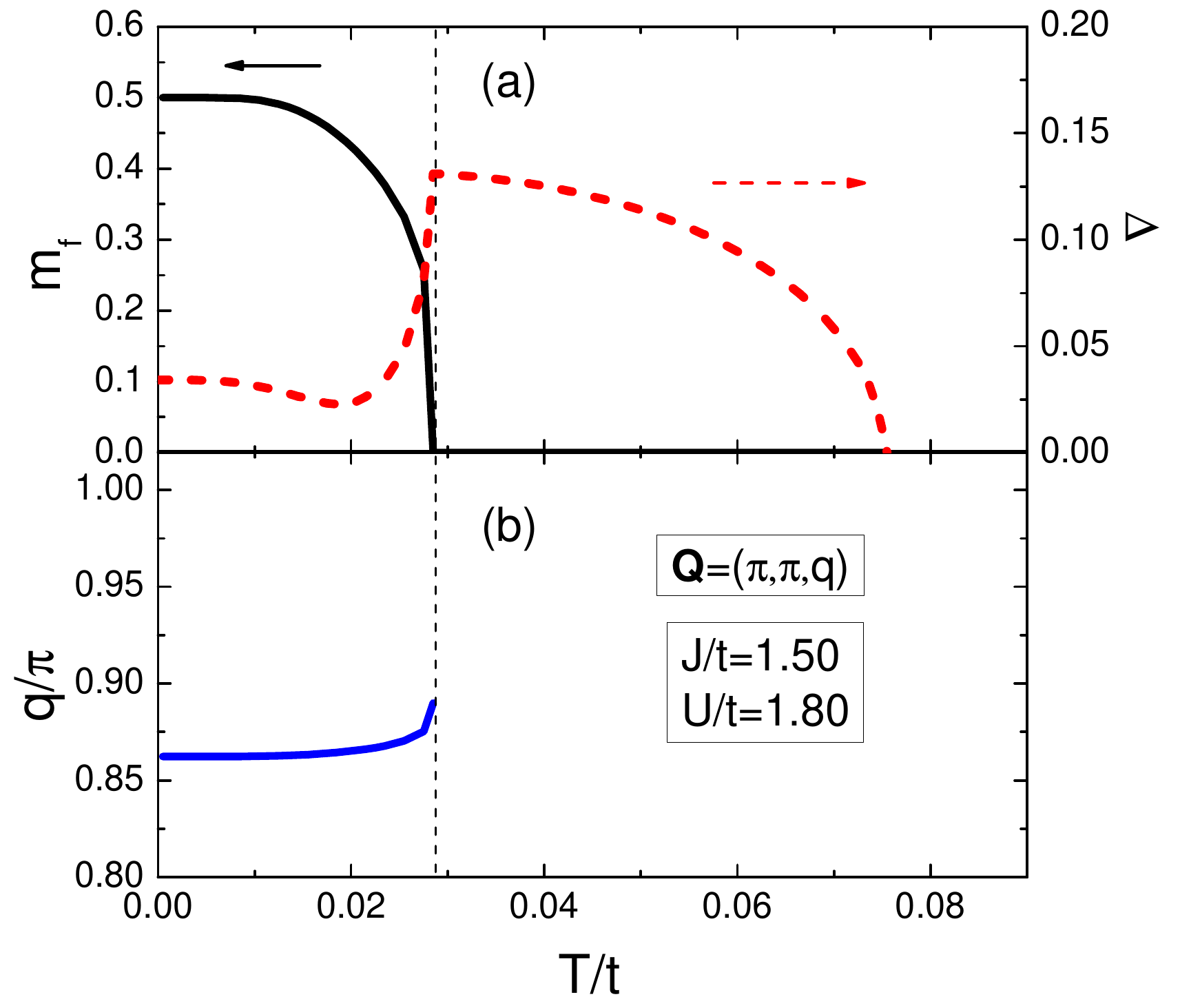}
\caption{(Color online) (a) Local moment magnetisation amplitude (left vertical scale) and superconducting gap (right vertical scale), and (b) the magnetic wave vector as a function of temperature for fixed $n_{c}=0.90$, $J/t=1.50$ and $U/t=1.80$.}
\label{order_temp1} 
\end{figure}

Figure \ref{order_temp1}(a) shows the dependence of $m_{f}$ and $\Delta$ with the temperature, for fixed $n_{c}=0.90$, $J/t=1.5$ and $U/t=1.8$.
The local-moment magnetisation amplitude decreases steadily with the temperature, while the superconducting gap is strongly influenced by the magnetic state. 
At low temperatures, superconductivity coexists with magnetism, though with a modest gap; when $m_f$ disappears, the gap increases considerably, and the system is purely superconducting until the temperature reaches $T_c$, at which $\Delta\to0$.  
Concerning the magnetic modes, the wave vector is hardly disturbed by the temperature, as shown in Fig.\,\ref{order_temp1}(b); 
only when the temperature approaches $T_{N}$, a small increase is noticeable in $q$.   

\begin{figure}[b]
\includegraphics[scale=0.30]{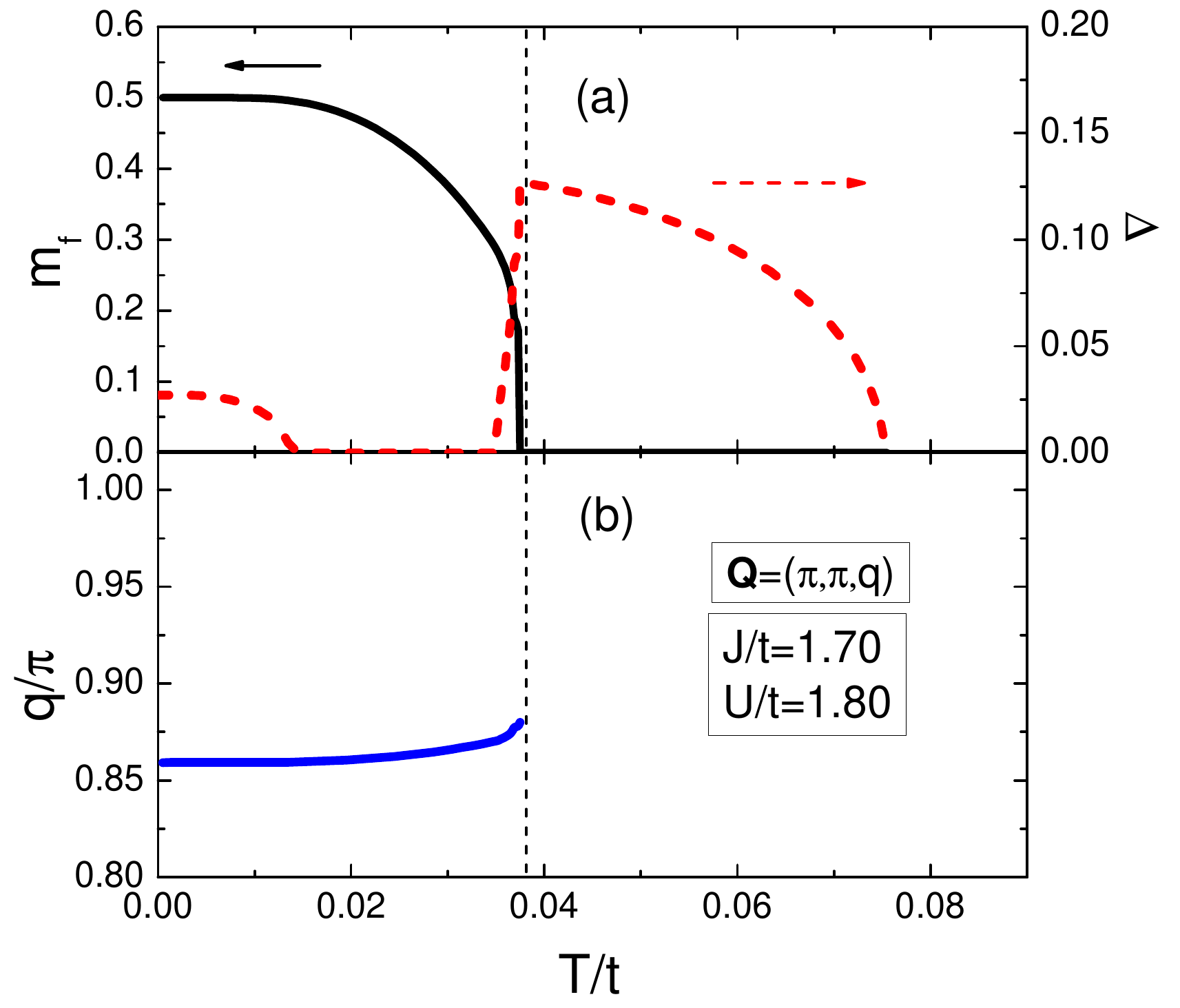}
\caption{(Colour online) Same as Fig.\,\ref{order_temp1}, but for $J/t=1.70$.}
\label{order_temp2} 
\end{figure}

Increasing $J/t$ favours magnetism, and one expects more pronounced effects in the behaviour of the gap with the temperature. 
Indeed, as displayed in Fig.\,\ref{order_temp2}(a), a slight increase in $J/t$ (in comparison with the value used in Fig.\,\ref{order_temp1}) results in a complete suppression of $\Delta$ within a range of temperatures.
However, when $T = T_{N}$ magnetism disappears, which in turn favours the reemergence of SC.
In addition, the behaviour of $q$ with $T/t$ depicted in Fig.\,\ref{order_temp2}(b) is very similar to that of Fig.\,\ref{order_temp1}(b). 
Therefore, while magnetism seems to be hardly affected by the coexistence with superconductivity, the converse is not true:   the behaviour of the SC order parameter is strongly dependent on whether or not there is magnetic order in the background.
We should mention that this re-entrant-like behaviour of the SC order parameter may be attributed to the absence of magnetic fluctuations for $T > T_{N}$ in the mean-field solution.

\begin{figure}[t]
\includegraphics[scale=0.31]{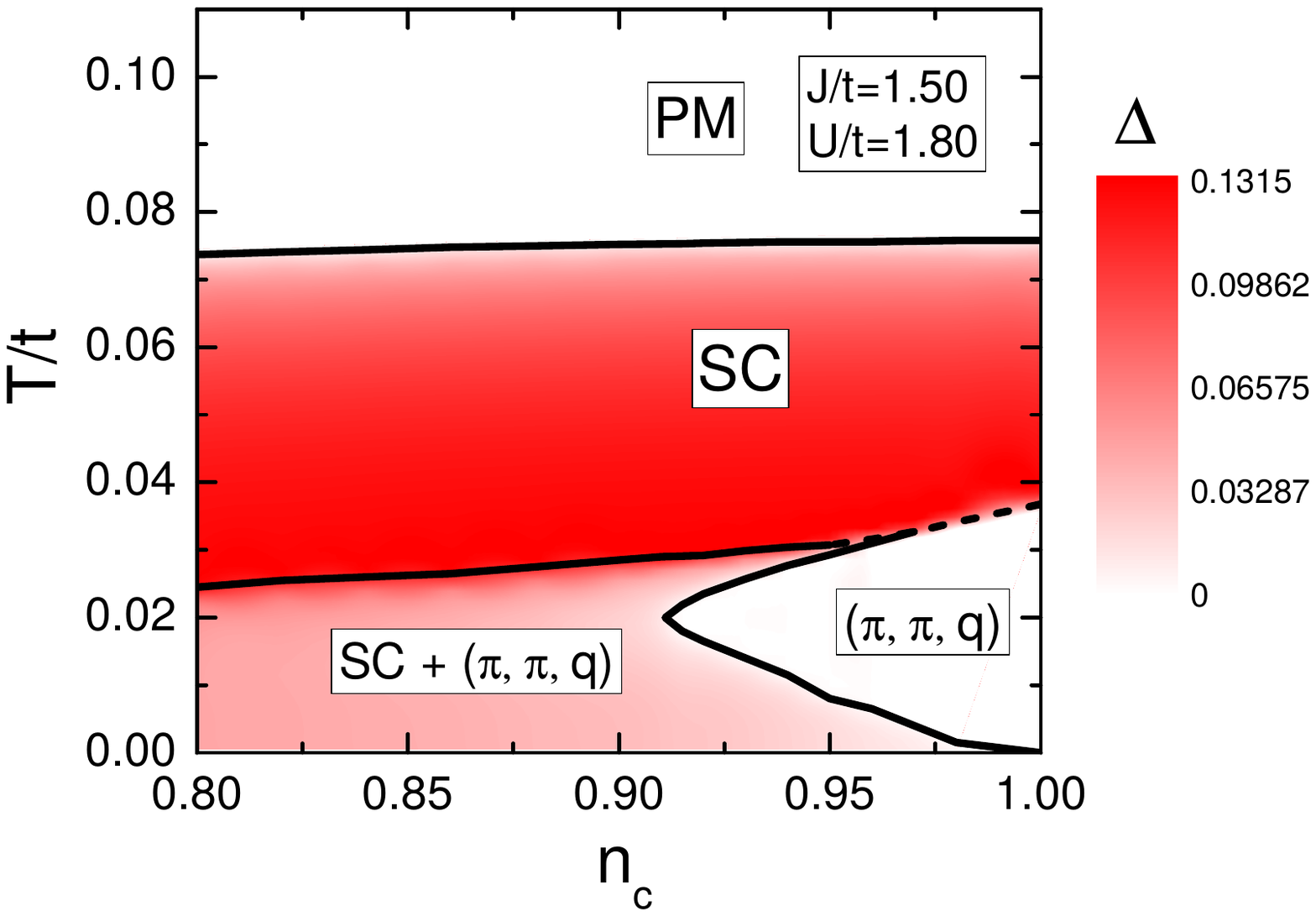}
\caption{(Colour online) Phase diagram of temperature as function of electronic density for $J/t=1.5$ and $U/t=1.8$.
The magnitude of the superconducting gap, $\Delta$, is mapped by the intensity of the shading in the SC regions; the numerical scale is shown to the right of the plotting area.}
\label{diag_temp_nc} 
\end{figure}

By carrying out similar analyses for other fillings while keeping fixed $J/t$ and $U/t$, we obtain phase diagrams like the one depicted in Fig.\,\ref{diag_temp_nc}.
For the choice of parameters in the figure, a reentrant superconducting phase is present for $ 0.91 \lesssim n_c \leq 1$, and one should have in mind that the decrease in the gap depicted in Fig.\,\ref{order_temp1}(a) does not configure as a reentrant behaviour.
We also note that the lobe around the reentrant SC region widens as $n_{c}$ approaches half filling; this can be attributed to the competition between SC and magnetism in the ground state which, as displayed in Fig.\,\ref{ncs}, is more detrimental to SC the closer one gets to half filling, and is enhanced as $J/t$ increases.
We have explicitly verified (not shown) that as $J/t$ increases, the lobe extends to lower electronic densities, while the boundary between the coexistence region with the purely SC one also rises. 
Again, this is due to magnetic order being progressively favoured as $J/t$ is increased, which increases the temperature interval during which SC is suppressed; also, larger $J/t$ allows the purely magnetic region to be extended to smaller fillings.

\begin{figure}[t]
\includegraphics[scale=0.3]{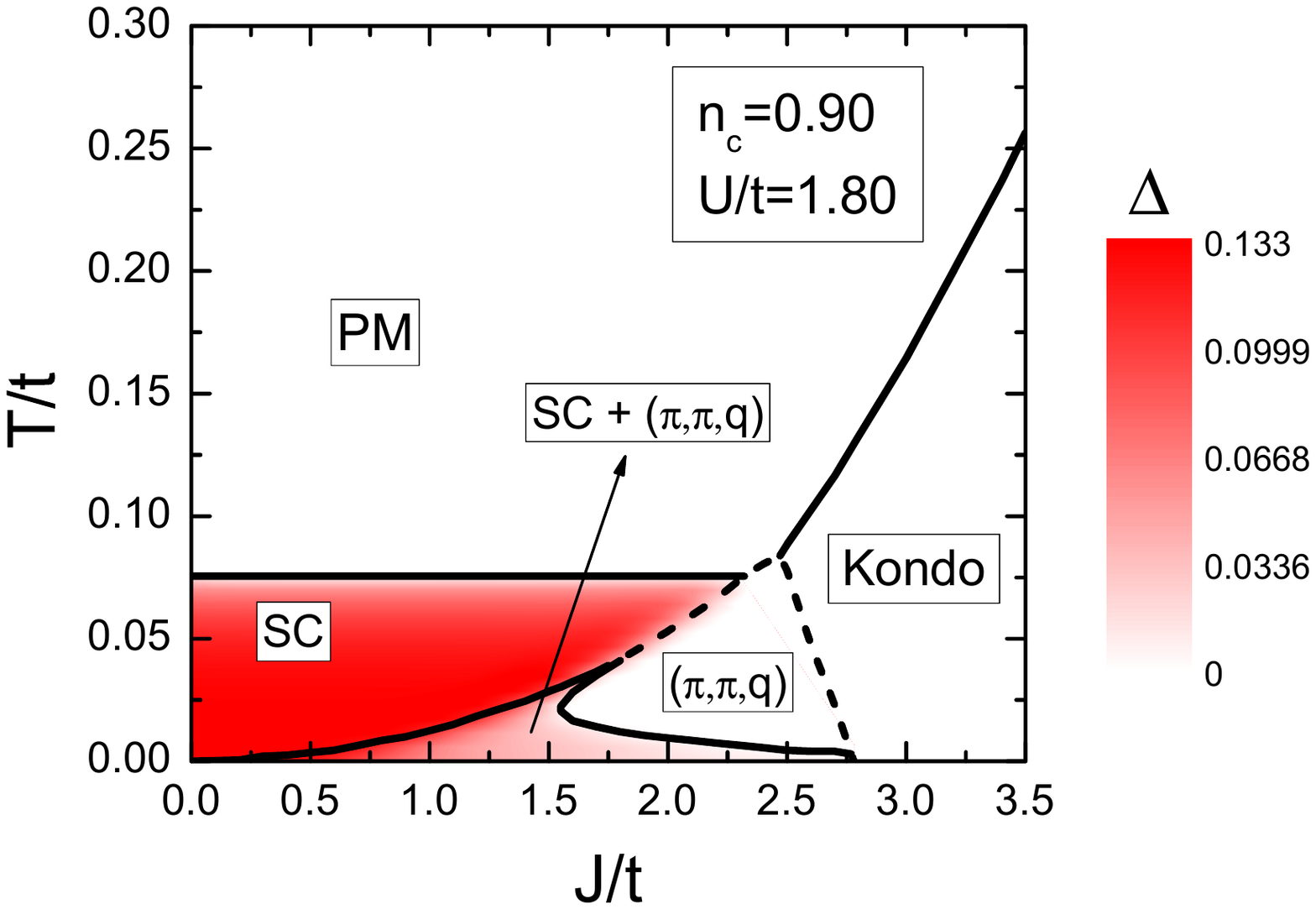}
\caption{(Colour online) Doniach-like phase diagram for fixed $n_{c}=0.90$ and $U/t=1.8$. The magnitude of the superconducting gap, $\Delta$, is mapped by the intensity of the shading in the SC regions; the numerical scale is shown to the right of the plotting area.}
\label{doniach} 
\end{figure}

Further insight into the main features of the multi-dimensional phase diagram can be acquired by exploring a $T/t\times J/t$ section for fixed $n_c$ and $U/t$, such as the one shown in Fig.\,\ref{doniach}; when $U=0$ (KLM) this is known as the Doniach phase diagram.\cite{Doniach1977,Costa17a}
We note that the coexistence between SC and a spiral phase with $\mathbf{Q}=(\pi,\pi,q)$ is induced as $J/t$ increases. 
However, the balance is somewhat delicate, since if $J/t$ increases too much one reaches a non-SC spiral phase (also in the shape of a lobe, as in Fig.\,\ref{diag_temp_nc}), which eventually leads to a Kondo phase.
The boundary at which the spiral phase disappears is the N\'eel temperature, and it is interesting to see that the RKKY signature of the interaction between local moments, $T_N\propto J^2$, is not modified by the presence of the pairing interaction. 
Another interesting aspect is that the transition temperature to the  (normal) paramagnetic (PM) phase, $T_{c}$, is hardly dependent on $J/t$; the reason for this can be traced back to the fact that once $m_f$ vanishes, ceases the influence of the exchange $J/t$ on the SC state. 
That is, only the boundaries involving some intervening magnetic ordering are influenced by the exchange coupling.  

\begin{figure}[t]
\includegraphics[scale=0.3]{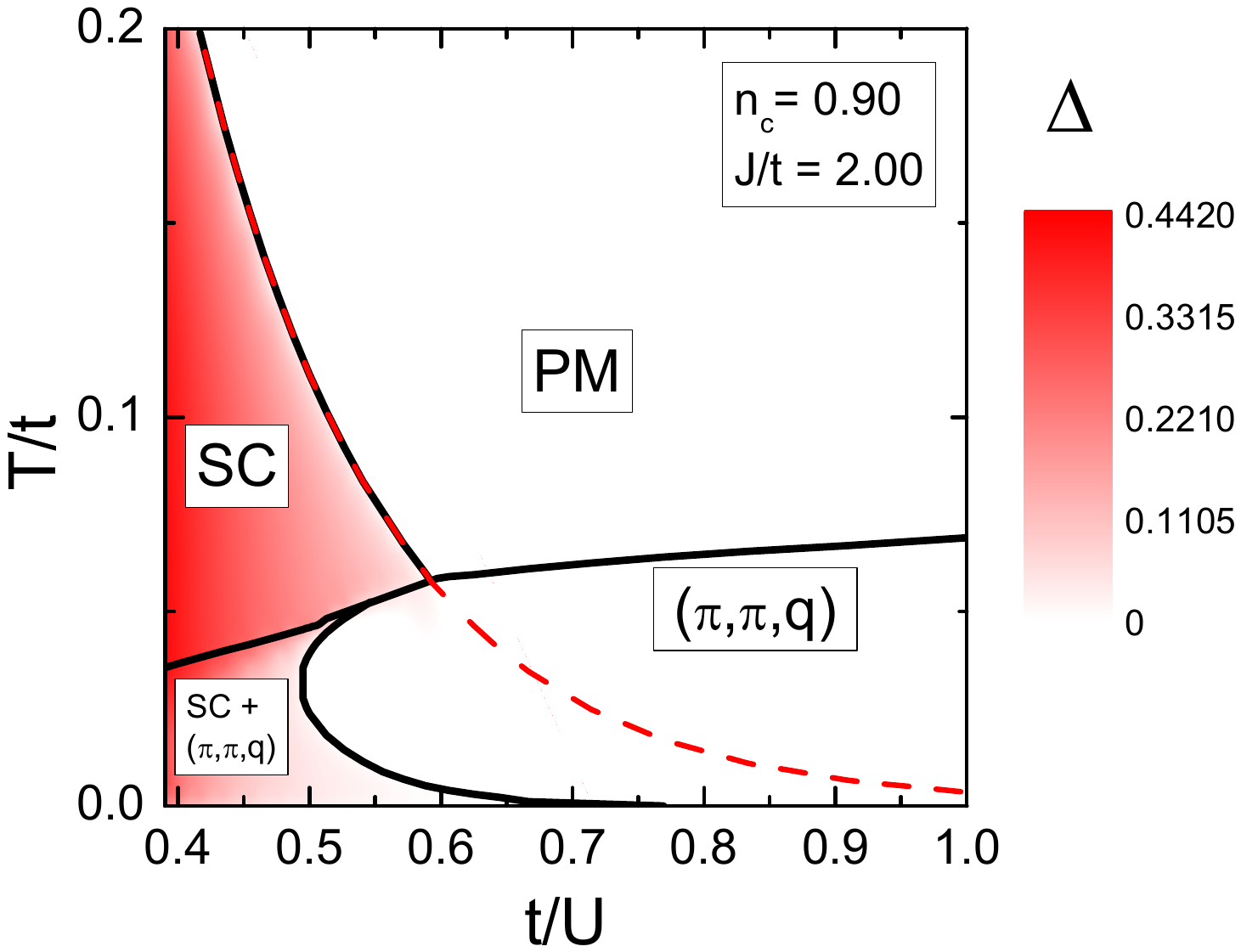}
\caption{(Colour online) 
Temperature vs \emph{inverse} on-site attraction, for fixed $n_c$ and $J/t$.  
The dashed (red) line is the superconducting critical temperature, $T_c$, for the attractive Hubbard model (i.e., $J=0$), which decreases exponentially with $t/U$, The nearly straight line across the plot is the N\'eel temperature, $T_N$. 
The magnitude of the superconducting gap, $\Delta$, is mapped by the intensity of the shading in the SC regions; the numerical scale is shown to the right of the plotting area.
}
\label{fig:TvsU} 
\end{figure}

We now turn to discussing a section $T/t\times U/t$ of the phase diagram, but plotted in terms of $t/U$. 
Figure \ref{fig:TvsU} shows data for fixed $n_c$ and $J/t$. 
The dashed (red) line is the phase boundary for the $J=0$ case, highlighting the usual BCS exponential dependence of $T_c$ with $t/U$, similar to the right-hand side of Eq.\,\eqref{eq:Delta}.
The effect of the Kondo coupling is apparent from Fig.\,\ref{fig:TvsU}: 
First, instead of the exponential decrease of $T_c(U)$ for large $t/U$, here we see that superconductivity is suppressed more abruptly: it only exists in the ground state up to $t/U_c\approx 0.77$ (see also Fig.\,\ref{diag090}), and it is also limited to a finite range of $t/U$ at finite temperatures.
Second, a spiral magnetic phase emerges at low temperatures, which is completely absent when $J=0$, and its N\'eel temperature, $T_N(U)$, increases almost linearly with increasing $t/U$; also, coexistence between magnetic and superconductor orderings develops at lower temperatures. 
As noted before, when the spiral phase vanishes at higher temperatures, the SC-normal boundary is the same as that when $J=0$.  
Third, a reentrant SC+spiral phase appears around $t/U_c$, whose size varies with $J/t$ and $n_c$.
In summary, for given $J/t$ and $n_c$, superconductivity is suppressed by magnetism for $U\lesssim U_c(n_c,J)$, and by thermal fluctuations [at some $T_c(U,J,n_c)$] when $U\gtrsim U_c(n_c,J)$.

\section{Discussion}\label{sec:discussion}

Given that magnetism in our model originates from local moments, it is instructive to make a rough comparison of our results with the experimental findings for QBC, mostly for the $R$Ni$_2$B$_2$C family of compounds. 
We first consider the sequence of the magnetic rare earths, namely $R=$ Lu, Tm, Er, Ho, Dy, Tb, and Gd, grouped together according to the de Gennes factor \cite{Canfield98,Muller2001,Gupta2006}.
The latter is defined as $dG\equiv  (g-1)^2j(j+1)$, with $j(j+1)$ being the eigenvalue of total angular momentum, and $g$ the Land\'e factor; experimentally, $T_N$ scales with $dG$ \cite{Canfield98,Muller2001,Gupta2006}.
Although our model does not take into account orbital angular momentum, it is tempting to replace the `bare' Kondo exchange coupling $J$ by an effective one, $J\cdot \sqrt{dG}$, so that $T_N \propto J^2$, as displayed in Fig.\,\ref{doniach}.
With this replacement we see that for $R=$ Lu ($J=0$ in our approach) the material is a non-magnetic superconductor, while for large $dG$ (or large $J$ in our approach), the compound is magnetic, but without SC. 
In between these two rare earths, magnetism and SC coexist and compete with each other.

It is also worth mentioning the similarities and differences between our findings and those of Ref.\,\cite{Karmakar16}, which considers the case of a ferromagnetic Kondo coupling.
In both cases one finds a multitude of magnetic phases coexisting, or not, with superconductivity in the ground state for a range of parameters.
Other features, such as a first order transition at half filling and the formation of a tail in the SC order parameter away from half filling (our Figs.\,\ref{order_100} and \ref{Deltas}, respectively) are also observed in Ref.\,\cite{Karmakar16}; 
here we attribute the appearance of tails to the metallic character of the normal state [see the discussion following Eq.\,\eqref{eq:Delta}].  
On the other hand, the main differences in the results are the absence of SC slightly away from half filling even for large $U/t$, which we do not observe here, and the absence of a singlet phase at large $J/t$, which we obtain here.
These differences may be attributed to the ferromagnetic interaction considered in Ref.\,\cite{Karmakar16}. 
Most importantly, our work goes beyond ground state properties.
We provide an extensive investigation of finite temperature properties, revealing the behaviour of several quantities with temperature, including that of the magnetic wave vector. 
In particular, we highlight the thermal effects brought about by the competition between superconductivity and magnetism.

\section{\label{sec:concl} Conclusions}
  
We have investigated the issue of coexistence between localised magnetism and superconductivity within an effective microscopic model, comprising of local moments on every site, coupled to conduction electrons by a Kondo-like interaction, $J$; pairing is made possible by an attractive on-site interaction, $U$, between the conduction electrons, whose density is $n_c$.
The model is solved on a simple-cubic lattice through a Hartree-Fock approximation within a semi-classical implementation \cite{Costa17a}.

We have first mapped out the ground state phases in the parameter space $(n_c,J,U)$. 
For a fixed band filling, the overall features of the $U\ vs.\ J$ phase diagrams can be summarized as follows: a Kondo (singlet) phase stabilizes at large $J$, and magnetic and superconducting phases in the smaller $J$ region; in the latter, as $U$ increases, the purely spiral phase at small $U$ evolves to a phase with superconducting coexisting with spiral antiferromagnetic (SAFM) arrangements (the magnetic wave vector has a weak dependence with $U$ in this phase), and then to a superconducting phase coexisting with a N\'eel phase. 
Therefore, the stronger the on-site attraction is, the greater is the tendency towards N\'eel order in the coexistence phase.  
Also, at half filling the transition is from a superconducting state to an insulating state, and no spiral phases are stabilized. 
For completeness, we note that we have not carried out a systematic investigation of the low electronic density region, to map out the boundaries of the ferromagnetic (FM) state, due to the plethora of different magnetic states that emerge in this regime \cite{Costa17a}.

Thermal fluctuations also play an important role in suppressing the ordered phases. 
When the ground state displays coexistence SC+SAFM, increasing the temperature induces two successive transitions: first, the SAFM phase is suppressed, while the SC phase is suppressed at higher temperatures; we note here that the magnetic wave vector shows a weak dependence with the temperature within a phase with SAFM, but it turns stronger just before the SAFM phase is suppressed. 
When the Kondo phase is the ground state, increasing thermal fluctuations breaks the singlets, and the system becomes paramagnetic at sufficiently high temperatures.   
Interestingly, in some instances increasing the temperature induces reentrant behaviour, namely SC+SAFM $\to$ SAFM $\to$  SC+SAFM $\to$ SC $\to$ PM.

Our model reproduces some experimental facts about the borocarbides.
First, when the bare exchange coupling $J$ is dressed by the rare earths' total angular momentum, the N\'eel temperature increases linearly with the de Gennes factor, $dG$, as observed experimentally.
Second, if we correlate the inverse on-site attraction with the ionic radius, we obtain a satisfactory description of the trend in magnetic and superconducting phases observed in $R$Ni$_2$B$_2$C, with $R$ running across all rare earths; this analysis highlights the difference between model parameters and the `real' parameters. 
The multitude of spiral magnetic phases we found by varying electron count and coupling constants is also a trademark of the borocarbides family.

In conclusion, our results indicate that the Kondo--attractive-Hubbard model provides a unifying view of the competition between localised magnetism and superconductivity. 

\begin{acknowledgments}

The authors are grateful to the Brazilian Agencies CAPES, CNPq and FAPERJ for financial support.

\end{acknowledgments}


\appendix*
\section{Mean-field approximation}
\label{Ap}

The Hartree-Fock approximation on the attractive potential term, Eq.\,\eqref{hamilHub}, leads to
\begin{align}
\nonumber
\mathcal{H}_{U_{(MF)}} = \sum_{i}\bigg[  &- U \frac{ n_{c}}{2} \big( c^{\dagger}_{i \uparrow} c_{i \uparrow} + c^{\dagger}_{i \downarrow} c_{i \downarrow} \big) + \frac{U n^{2}_{c}}{4} \\
\nonumber
& - t\Delta \big( c^{\dagger}_{i \uparrow} c^{\dagger}_{i \downarrow} + c_{i \downarrow} c_{i \uparrow} \big) + \frac{(t\Delta)^2}{U}\\
& + 2 U \langle \mathbf{s}^{c}_{i} \rangle\! \cdot\! \mathbf{s}^{c}_{i} - U \langle \mathbf{s}^{c}_{i} \rangle\! \cdot\!\langle \mathbf{s}^{c}_{i} \rangle \bigg] ,
\label{hamilHub2}
\end{align}
with 
\begin{equation}\label{eq:deltaU}
	\frac{t\Delta}{U} = \langle c^{\dagger}_{i \uparrow} c^{\dagger}_{i \downarrow} \rangle 
				= \langle c_{i \downarrow} c_{i \uparrow} \rangle,
\end{equation}	
where $\Delta$ is the
(dimensionless)
order parameter of superconductivity, and $n_{c} = \langle c^{\dagger}_{i \uparrow} c_{i \uparrow}  + c^{\dagger}_{i \downarrow} c_{i \downarrow} \rangle $ is the electronic density of conduction electrons, both taken as homogeneous throughout the sites. The operator $ \mathbf{s}^{c}_{i} $ is defined in Eq.\eqref{Sc}.
The inclusion of the hopping integral $t$ in Eqs.\,\eqref{hamilHub2} and \eqref{eq:deltaU} is solely to render $\Delta$ dimensionless.

Following the procedure presented in Ref.\,\onlinecite{Costa17a}, the Kondo lattice term, Eq.\,\eqref{hamilKLM}, can be decoupled as
\begin{widetext}
\begin{align} 
\nonumber \mathcal{H}_{K_{(MF)}} = & -t \sum_{\langle i,j \rangle, \sigma} \big( c^{\dagger}_{i \sigma} c^{\phantom{\dagger}}_{j \sigma} + \mathrm{H.c.} \big) 
+ J \sum_{i} \big( \mathbf{S}_{i}\! \cdot\! \langle \mathbf{s}^{c}_{i} \rangle + \langle \mathbf{S}_{i} \rangle \!\cdot\! \mathbf{s}^{c}_{i} \big) 
+ \frac{J}{2} \sum_{i} \big( \mathbf{V}^{c}_{i}\! \cdot\! \langle \mathbf{V}^{f}_{i} \rangle + \langle \mathbf{V}^{c}_{i} \rangle \!\cdot\! \mathbf{V}^{f}_{i} \big) \\ 
&- \frac{3 J}{2} \sum_{i} \big( V^{0}_{i c} \langle V^{0}_{i f} \rangle + \langle V^{0}_{i c} \rangle V^{0}_{i f} \big) - \frac{J}{2} \sum_{i} \langle \mathbf{V}^{c}_{i} \rangle\! \cdot\! \langle \mathbf{V}^{f}_{i} \rangle + \frac{3 J}{2} \sum_{i} \langle V^{0}_{i c} \rangle \langle V^{0}_{i f} \rangle
- J \sum_{i} \langle \mathbf{S}_{i} \rangle\! \cdot\! \langle \mathbf{s}^{c}_{i} \rangle,
\label{hamilKLM2}
\end{align}
\end{widetext}
with the operators
\begin{eqnarray}\label{singlet_hyb}
V^{0}_{i c}= {V^{0}_{i f}}^{\dagger} = \frac{1}{2}\sum_{\alpha, \beta = \pm} c^{\dagger}_{i \alpha} \mathbb{I}_{\alpha, \beta} f^{\phantom{\dagger}}_{i \beta},
\end{eqnarray}
and
\begin{eqnarray}\label{triplet_hyb}
\mathbf{V}_{i c} = \mathbf{V}^{\dagger}_{i f} = \frac{1}{2} \sum_{\alpha, \beta = \pm} c^{\dagger}_{i \alpha} \boldsymbol{\sigma}_{\alpha, \beta} f^{\phantom{\dagger}}_{i \beta},
\end{eqnarray}
being singlet and triplet hybridization terms, respectively.

Within a spiral magnetic approach, the mean values $\langle \mathbf{S}_{i} \rangle$ and $\langle \mathbf{s}^{c}_{i} \rangle$ are taken as 
\begin{align}\label{Si}
\langle \mathbf{S}_{i} \rangle = m_{f} \big[ \cos \left(\mathbf{Q}\!\cdot\! \mathbf{R}_{i}\right), \sin \left(\mathbf{Q}\!\cdot\! \mathbf{R}_{i}\right), 0 \big]
\end{align}
and
\begin{align}\label{Sci}
\langle \mathbf{s}^{c}_{i} \rangle = -m_{c} \big[ \cos \left(\mathbf{Q}\!\cdot\! \mathbf{R}_{i}\right), \sin \left(\mathbf{Q}\!\cdot\! \mathbf{R}_{i}\right),0 \big],
\end{align}
with 
\begin{align} 
\mathbf{Q}=(q_{x}, q_{y}, q_{z})
\label{Q}
\end{align}
being the magnetic wave vector, and $\mathbf{R}_{i}$ the position vector of site \textit{i} on the lattice.

By the same token, the mean values of the hybridization operators are chosen as
\begin{equation}\label{singlet_hyb_mean_v}
\langle V^{0}_{i c} \rangle = \langle {V^{0}_{i f}}^{\dagger} \rangle = -V
\end{equation}
and
\begin{equation}\label{triplet_hyb_mean_v}
\langle \mathbf{V}_{i c} \rangle  = \langle  {\mathbf{V}_{i f}}^{\dagger} \rangle = V' \big[ \cos \left(\mathbf{Q}\!\cdot\! \mathbf{R}_{i}\right), \sin \left(\mathbf{Q}\!\cdot\! \mathbf{R}_{i}\right),0 \big].
\end{equation}

Then, our mean-field Hamiltonian, Eq.\,\eqref{hamilKaH}, is obtained by substituting Eqs.\,\eqref{hamilHub2} and \eqref{hamilKLM2} in Eq.\,\eqref{hamil}, using Eqs.\,\eqref{Si}, \eqref{Sci}, \eqref{singlet_hyb_mean_v} and \eqref{triplet_hyb_mean_v} for the mean values of the spin and hybridization operators, and performing a discrete Fourier transform, with periodic boundary conditions.
Such Hamiltonian is two-fold degenerate when written in the spinor Nambu basis $$ \mathbf{\Psi}^{\dagger}_{\mathbf{k}} = 
\big( 
c^{\dagger}_{\mathbf{k} \uparrow},
c^{\dagger}_{\mathbf{k} + \mathbf{Q} \downarrow},
f^{\dagger}_{\mathbf{k} \uparrow}, f^{\dagger}_{\mathbf{k} + \mathbf{Q} \downarrow},
c_{-\mathbf{k} \downarrow},
c_{-\mathbf{k} -\mathbf{Q} \uparrow},
f_{-\mathbf{k} \downarrow},
f_{-\mathbf{k} -\mathbf{Q} \uparrow}
\big), $$ 
leading to
\begin{align}
\mathcal{H}_{MF} = \Psi^{\dagger}_{\mathbf{k}} \hat{H} \Psi_{\mathbf{k}} + const.,
\end{align}
with
\begin{widetext}
\begin{equation}
\label{Matrix_hamil}
\hat{H}=
\left(
    {
    \begin{array}{cccccccc}
\epsilon_{\mathbf{k}} - \tilde{\mu} & \frac{1}{2}J m_{f} - U m_{c} & \frac{3}{4}J V & \frac{1}{4} J V' 
& -t\Delta & 0 & 0 & 0 \\
\\
\frac{1}{2}J m_{f} - U m_{c} & \epsilon_{\mathbf{k} + \mathbf{Q}} - \tilde{\mu} & \frac{1}{4} J V' & \frac{3}{4}J V
& 0 & t\Delta & 0 & 0  \\
\\
\frac{3}{4}J V & \frac{1}{4} J V' & \epsilon_{f} & -\frac{1}{2}J m_{c}
& 0 & 0 & 0 & 0 \\
\\
\frac{1}{4} J V' & \frac{3}{4}J V & -\frac{1}{2}J m_{c} & \epsilon_{f}
& 0 & 0 & 0 & 0 \\
\\
-t\Delta & 0 & 0 & 0
& -(\epsilon_{-\mathbf{k}} - \tilde{\mu}) & -(\frac{1}{2}J m_{f} - U m_{c}) & -\frac{3}{4}J V & -\frac{1}{4} J V' \\
\\
0 & t\Delta & 0 & 0
& -(\frac{1}{2}J m_{f} - U m_{c}) & -(\epsilon_{-\mathbf{k}-\mathbf{Q}} - \tilde{\mu}) & -\frac{1}{4} J V' & -\frac{3}{4}J V \\
\\
0 & 0 & 0 & 0
& -\frac{3}{4}J V & -\frac{1}{4} J V' & -\epsilon_{f} & \frac{1}{2}J m_{c} \\
\\
0 & 0 & 0 & 0
& -\frac{1}{4} J V' & -\frac{3}{4}J V & \frac{1}{2}J m_{c} & -\epsilon_{f} \\
    \end{array}
    }
  \right).
\end{equation}
and
\begin{align}
const.= 2 N \bigg[ \mu \big(n_{c} - 1 \big) - \epsilon_{f} \big(n_{f} - 1 \big) + J m_{c} m_{f} + \frac{3}{2} J V^{2} - \frac{1}{2} J V'^{2} + \frac{U n_{c}^{2}}{4} + \frac{(t\Delta)^2}{U} - U m_{c}^{2} - \frac{U n_{c}}{2} \bigg].
\end{align}
\end{widetext}
Here, we defined $\tilde{\mu}=\big( \mu - \frac{1}{2} U n_{c} \big)$, in which $\mu$ and $\epsilon_{f}$ are included as Lagrange multipliers.

The eigenvalues of the Hamiltonian of Eq.\,\eqref{Matrix_hamil} are used to obtain the Helmholtz free energy, Eq.\,\eqref{Helmholtz}, and, consequently, the set of nonlinear equations, Eq.\,\eqref{selfcon}.

\bibliography{Costa_KaH}

\begin{thebibliography}{58}%
\makeatletter
\providecommand \@ifxundefined [1]{%
 \@ifx{#1\undefined}
}%
\providecommand \@ifnum [1]{%
 \ifnum #1\expandafter \@firstoftwo
 \else \expandafter \@secondoftwo
 \fi
}%
\providecommand \@ifx [1]{%
 \ifx #1\expandafter \@firstoftwo
 \else \expandafter \@secondoftwo
 \fi
}%
\providecommand \natexlab [1]{#1}%
\providecommand \enquote  [1]{``#1''}%
\providecommand \bibnamefont  [1]{#1}%
\providecommand \bibfnamefont [1]{#1}%
\providecommand \citenamefont [1]{#1}%
\providecommand \href@noop [0]{\@secondoftwo}%
\providecommand \href [0]{\begingroup \@sanitize@url \@href}%
\providecommand \@href[1]{\@@startlink{#1}\@@href}%
\providecommand \@@href[1]{\endgroup#1\@@endlink}%
\providecommand \@sanitize@url [0]{\catcode `\\12\catcode `\$12\catcode
  `\&12\catcode `\#12\catcode `\^12\catcode `\_12\catcode `\%12\relax}%
\providecommand \@@startlink[1]{}%
\providecommand \@@endlink[0]{}%
\providecommand \url  [0]{\begingroup\@sanitize@url \@url }%
\providecommand \@url [1]{\endgroup\@href {#1}{\urlprefix }}%
\providecommand \urlprefix  [0]{URL }%
\providecommand \Eprint [0]{\href }%
\providecommand \doibase [0]{http://dx.doi.org/}%
\providecommand \selectlanguage [0]{\@gobble}%
\providecommand \bibinfo  [0]{\@secondoftwo}%
\providecommand \bibfield  [0]{\@secondoftwo}%
\providecommand \translation [1]{[#1]}%
\providecommand \BibitemOpen [0]{}%
\providecommand \bibitemStop [0]{}%
\providecommand \bibitemNoStop [0]{.\EOS\space}%
\providecommand \EOS [0]{\spacefactor3000\relax}%
\providecommand \BibitemShut  [1]{\csname bibitem#1\endcsname}%
\let\auto@bib@innerbib\@empty
\bibitem [{\citenamefont {Mukuda}\ \emph {et~al.}(2012)\citenamefont {Mukuda},
  \citenamefont {Shimizu}, \citenamefont {Iyo},\ and\ \citenamefont
  {Kitaoka}}]{Mukuda12}%
  \BibitemOpen
  \bibfield  {author} {\bibinfo {author} {\bibfnamefont {H.}~\bibnamefont
  {Mukuda}}, \bibinfo {author} {\bibfnamefont {S.}~\bibnamefont {Shimizu}},
  \bibinfo {author} {\bibfnamefont {A.}~\bibnamefont {Iyo}}, \ and\ \bibinfo
  {author} {\bibfnamefont {Y.}~\bibnamefont {Kitaoka}},\ }\href {\doibase
  10.1143/JPSJ.81.011008} {\bibfield  {journal} {\bibinfo  {journal} {J. Phys.
  Soc. Jpn.}\ }\textbf {\bibinfo {volume} {81}},\ \bibinfo {pages} {011008}
  (\bibinfo {year} {2012})}\BibitemShut {NoStop}%
\bibitem [{\citenamefont {Scalapino}(2012)}]{Scalapino12}%
  \BibitemOpen
  \bibfield  {author} {\bibinfo {author} {\bibfnamefont {D.~J.}\ \bibnamefont
  {Scalapino}},\ }\href {\doibase 10.1103/RevModPhys.84.1383} {\bibfield
  {journal} {\bibinfo  {journal} {Rev. Mod. Phys.}\ }\textbf {\bibinfo {volume}
  {84}},\ \bibinfo {pages} {1383} (\bibinfo {year} {2012})}\BibitemShut
  {NoStop}%
\bibitem [{\citenamefont {Stewart}(2011)}]{Stewart11}%
  \BibitemOpen
  \bibfield  {author} {\bibinfo {author} {\bibfnamefont {G.~R.}\ \bibnamefont
  {Stewart}},\ }\href {\doibase 10.1103/RevModPhys.83.1589} {\bibfield
  {journal} {\bibinfo  {journal} {Rev. Mod. Phys.}\ }\textbf {\bibinfo {volume}
  {83}},\ \bibinfo {pages} {1589} (\bibinfo {year} {2011})}\BibitemShut
  {NoStop}%
\bibitem [{\citenamefont {Si}\ \emph {et~al.}(2016)\citenamefont {Si},
  \citenamefont {Yu},\ and\ \citenamefont {Abrahams}}]{Si16}%
  \BibitemOpen
  \bibfield  {author} {\bibinfo {author} {\bibfnamefont {Q.}~\bibnamefont
  {Si}}, \bibinfo {author} {\bibfnamefont {R.}~\bibnamefont {Yu}}, \ and\
  \bibinfo {author} {\bibfnamefont {E.}~\bibnamefont {Abrahams}},\ }\href
  {http://dx.doi.org/10.1038/natrevmats.2016.17} {\bibfield  {journal}
  {\bibinfo  {journal} {Nat. Rev. Mat.}\ }\textbf {\bibinfo {volume} {1}},\
  \bibinfo {pages} {16017 EP } (\bibinfo {year} {2016})}\BibitemShut {NoStop}%
\bibitem [{\citenamefont {Pfleiderer}(2009)}]{Pfleiderer09}%
  \BibitemOpen
  \bibfield  {author} {\bibinfo {author} {\bibfnamefont {C.}~\bibnamefont
  {Pfleiderer}},\ }\href {\doibase 10.1103/RevModPhys.81.1551} {\bibfield
  {journal} {\bibinfo  {journal} {Rev. Mod. Phys.}\ }\textbf {\bibinfo {volume}
  {81}},\ \bibinfo {pages} {1551} (\bibinfo {year} {2009})}\BibitemShut
  {NoStop}%
\bibitem [{\citenamefont {Weng}\ \emph {et~al.}(2016)\citenamefont {Weng},
  \citenamefont {Smidman}, \citenamefont {Jiao}, \citenamefont {Lu},\ and\
  \citenamefont {Yuan}}]{Weng16}%
  \BibitemOpen
  \bibfield  {author} {\bibinfo {author} {\bibfnamefont {Z.~F.}\ \bibnamefont
  {Weng}}, \bibinfo {author} {\bibfnamefont {M.}~\bibnamefont {Smidman}},
  \bibinfo {author} {\bibfnamefont {L.}~\bibnamefont {Jiao}}, \bibinfo {author}
  {\bibfnamefont {X.}~\bibnamefont {Lu}}, \ and\ \bibinfo {author}
  {\bibfnamefont {H.~Q.}\ \bibnamefont {Yuan}},\ }\href
  {http://stacks.iop.org/0034-4885/79/i=9/a=094503} {\bibfield  {journal}
  {\bibinfo  {journal} {Rep. Prog. Phys.}\ }\textbf {\bibinfo {volume} {79}},\
  \bibinfo {pages} {094503} (\bibinfo {year} {2016})}\BibitemShut {NoStop}%
\bibitem [{\citenamefont {Steglich}\ and\ \citenamefont
  {Wirth}(2016)}]{Steglich16}%
  \BibitemOpen
  \bibfield  {author} {\bibinfo {author} {\bibfnamefont {F.}~\bibnamefont
  {Steglich}}\ and\ \bibinfo {author} {\bibfnamefont {S.}~\bibnamefont
  {Wirth}},\ }\href {http://stacks.iop.org/0034-4885/79/i=8/a=084502}
  {\bibfield  {journal} {\bibinfo  {journal} {Rep. Prog. Phys.}\ }\textbf
  {\bibinfo {volume} {79}},\ \bibinfo {pages} {084502} (\bibinfo {year}
  {2016})}\BibitemShut {NoStop}%
\bibitem [{\citenamefont {Cava}\ \emph {et~al.}(1994)\citenamefont {Cava},
  \citenamefont {Takagi}, \citenamefont {Zandbergen}, \citenamefont
  {Krajewski}, \citenamefont {\surname{Peck Jr.}}, \citenamefont {Sigerist},
  \citenamefont {Batlogg}, \citenamefont {\surname{Van Dover}}, \citenamefont
  {Felder}, \citenamefont {Mizuhashi}, \citenamefont {Lee}, \citenamefont
  {Eisaki},\ and\ \citenamefont {Uchida}}]{Cava94}%
  \BibitemOpen
  \bibfield  {author} {\bibinfo {author} {\bibfnamefont {R.~J.}\ \bibnamefont
  {Cava}}, \bibinfo {author} {\bibfnamefont {H.}~\bibnamefont {Takagi}},
  \bibinfo {author} {\bibfnamefont {H.~W.}\ \bibnamefont {Zandbergen}},
  \bibinfo {author} {\bibfnamefont {J.~J.}\ \bibnamefont {Krajewski}}, \bibinfo
  {author} {\bibfnamefont {W.~F.}\ \bibnamefont {\surname{Peck Jr.}}}, \bibinfo
  {author} {\bibfnamefont {T.}~\bibnamefont {Sigerist}}, \bibinfo {author}
  {\bibfnamefont {B.}~\bibnamefont {Batlogg}}, \bibinfo {author} {\bibfnamefont
  {R.~B.}\ \bibnamefont {\surname{Van Dover}}}, \bibinfo {author}
  {\bibfnamefont {R.~J.}\ \bibnamefont {Felder}}, \bibinfo {author}
  {\bibfnamefont {K.}~\bibnamefont {Mizuhashi}}, \bibinfo {author}
  {\bibfnamefont {J.~O.}\ \bibnamefont {Lee}}, \bibinfo {author} {\bibfnamefont
  {H.}~\bibnamefont {Eisaki}}, \ and\ \bibinfo {author} {\bibfnamefont
  {S.}~\bibnamefont {Uchida}},\ }\href@noop {} {\bibfield  {journal} {\bibinfo
  {journal} {Nature}\ }\textbf {\bibinfo {volume} {367}},\ \bibinfo {pages}
  {252} (\bibinfo {year} {1994})}\BibitemShut {NoStop}%
\bibitem [{\citenamefont {Canfield}\ \emph {et~al.}(1998)\citenamefont
  {Canfield}, \citenamefont {Gammel},\ and\ \citenamefont
  {Bishop}}]{Canfield98}%
  \BibitemOpen
  \bibfield  {author} {\bibinfo {author} {\bibfnamefont {P.~C.}\ \bibnamefont
  {Canfield}}, \bibinfo {author} {\bibfnamefont {P.~L.}\ \bibnamefont
  {Gammel}}, \ and\ \bibinfo {author} {\bibfnamefont {D.~J.}\ \bibnamefont
  {Bishop}},\ }\href {\doibase http://dx.doi.org/10.1063/1.882396} {\bibfield
  {journal} {\bibinfo  {journal} {Phys. Today}\ }\textbf {\bibinfo {volume}
  {51}},\ \bibinfo {pages} {40} (\bibinfo {year} {1998})}\BibitemShut {NoStop}%
\bibitem [{\citenamefont {M\"uller}\ and\ \citenamefont
  {Narozhnyi}(2001)}]{Muller2001}%
  \BibitemOpen
  \bibfield  {author} {\bibinfo {author} {\bibfnamefont {K.-H.}\ \bibnamefont
  {M\"uller}}\ and\ \bibinfo {author} {\bibfnamefont {V.~N.}\ \bibnamefont
  {Narozhnyi}},\ }\href {http://stacks.iop.org/0034-4885/64/i=8/a=202}
  {\bibfield  {journal} {\bibinfo  {journal} {Rep. Prog. Phys.}\ }\textbf
  {\bibinfo {volume} {64}},\ \bibinfo {pages} {943} (\bibinfo {year}
  {2001})}\BibitemShut {NoStop}%
\bibitem [{\citenamefont {M\"uller}\ \emph {et~al.}(2002)\citenamefont
  {M\"uller}, \citenamefont {Fuchs}, \citenamefont {Drechsler},\ and\
  \citenamefont {Narozhnyi}}]{Muller2002}%
  \BibitemOpen
  \bibfield  {author} {\bibinfo {author} {\bibfnamefont {K.-H.}\ \bibnamefont
  {M\"uller}}, \bibinfo {author} {\bibfnamefont {G.}~\bibnamefont {Fuchs}},
  \bibinfo {author} {\bibfnamefont {S.-L.}\ \bibnamefont {Drechsler}}, \ and\
  \bibinfo {author} {\bibfnamefont {V.}~\bibnamefont {Narozhnyi}},\ }in\ \href
  {\doibase http://dx.doi.org/10.1016/S1567-2719(09)60007-X} {\emph {\bibinfo
  {booktitle} {Handbook of Magnetic Materials}}},\ \bibinfo {series} {Handbook
  of Magnetic Materials}, Vol.~\bibinfo {volume} {14},\ \bibinfo {editor}
  {edited by\ \bibinfo {editor} {\bibfnamefont {K.}~\bibnamefont {Buschow}}}\
  (\bibinfo  {publisher} {Elsevier},\ \bibinfo {year} {2002})\ pp.\ \bibinfo
  {pages} {199 -- 305}\BibitemShut {NoStop}%
\bibitem [{\citenamefont {Gupta}(2006)}]{Gupta2006}%
  \BibitemOpen
  \bibfield  {author} {\bibinfo {author} {\bibfnamefont {L.~C.}\ \bibnamefont
  {Gupta}},\ }\href@noop {} {\bibfield  {journal} {\bibinfo  {journal} {Adv.
  Phys.}\ }\textbf {\bibinfo {volume} {55}},\ \bibinfo {pages} {691} (\bibinfo
  {year} {2006})}\BibitemShut {NoStop}%
\bibitem [{\citenamefont {Wolowiec}\ \emph {et~al.}(2015)\citenamefont
  {Wolowiec}, \citenamefont {White},\ and\ \citenamefont {Maple}}]{Wolowiec15}%
  \BibitemOpen
  \bibfield  {author} {\bibinfo {author} {\bibfnamefont {C.}~\bibnamefont
  {Wolowiec}}, \bibinfo {author} {\bibfnamefont {B.}~\bibnamefont {White}}, \
  and\ \bibinfo {author} {\bibfnamefont {M.}~\bibnamefont {Maple}},\ }\href
  {\doibase https://doi.org/10.1016/j.physc.2015.02.050} {\bibfield  {journal}
  {\bibinfo  {journal} {Physica C}\ }\textbf {\bibinfo {volume} {514}},\
  \bibinfo {pages} {113 } (\bibinfo {year} {2015})},\ \bibinfo {note}
  {superconducting Materials: Conventional, Unconventional and
  Undetermined}\BibitemShut {NoStop}%
\bibitem [{\citenamefont {Rodriguez}\ \emph {et~al.}(2011)\citenamefont
  {Rodriguez}, \citenamefont {Araujo},\ and\ \citenamefont
  {Sacramento}}]{Rodriguez11}%
  \BibitemOpen
  \bibfield  {author} {\bibinfo {author} {\bibfnamefont {J.~P.}\ \bibnamefont
  {Rodriguez}}, \bibinfo {author} {\bibfnamefont {M.~A.~N.}\ \bibnamefont
  {Araujo}}, \ and\ \bibinfo {author} {\bibfnamefont {P.~D.}\ \bibnamefont
  {Sacramento}},\ }\href {\doibase 10.1103/PhysRevB.84.224504} {\bibfield
  {journal} {\bibinfo  {journal} {Phys. Rev. B}\ }\textbf {\bibinfo {volume}
  {84}},\ \bibinfo {pages} {224504} (\bibinfo {year} {2011})}\BibitemShut
  {NoStop}%
\bibitem [{\citenamefont {J\ifmmode~\mbox{\k{e}}\else \k{e}\fi{}drak}\ and\
  \citenamefont {Spa\l{}ek}(2011)}]{Jedrak11}%
  \BibitemOpen
  \bibfield  {author} {\bibinfo {author} {\bibfnamefont {J.}~\bibnamefont
  {J\ifmmode~\mbox{\k{e}}\else \k{e}\fi{}drak}}\ and\ \bibinfo {author}
  {\bibfnamefont {J.}~\bibnamefont {Spa\l{}ek}},\ }\href {\doibase
  10.1103/PhysRevB.83.104512} {\bibfield  {journal} {\bibinfo  {journal} {Phys.
  Rev. B}\ }\textbf {\bibinfo {volume} {83}},\ \bibinfo {pages} {104512}
  (\bibinfo {year} {2011})}\BibitemShut {NoStop}%
\bibitem [{\citenamefont {Abram}\ \emph {et~al.}(2013)\citenamefont {Abram},
  \citenamefont {Kaczmarczyk}, \citenamefont {J\ifmmode~\mbox{\k{e}}\else
  \k{e}\fi{}drak},\ and\ \citenamefont {Spa\l{}ek}}]{Abram13}%
  \BibitemOpen
  \bibfield  {author} {\bibinfo {author} {\bibfnamefont {M.}~\bibnamefont
  {Abram}}, \bibinfo {author} {\bibfnamefont {J.}~\bibnamefont {Kaczmarczyk}},
  \bibinfo {author} {\bibfnamefont {J.}~\bibnamefont
  {J\ifmmode~\mbox{\k{e}}\else \k{e}\fi{}drak}}, \ and\ \bibinfo {author}
  {\bibfnamefont {J.}~\bibnamefont {Spa\l{}ek}},\ }\href {\doibase
  10.1103/PhysRevB.88.094502} {\bibfield  {journal} {\bibinfo  {journal} {Phys.
  Rev. B}\ }\textbf {\bibinfo {volume} {88}},\ \bibinfo {pages} {094502}
  (\bibinfo {year} {2013})}\BibitemShut {NoStop}%
\bibitem [{\citenamefont {Rodriguez}\ \emph {et~al.}(2014)\citenamefont
  {Rodriguez}, \citenamefont {Araujo},\ and\ \citenamefont
  {Sacramento}}]{Rodriguez14}%
  \BibitemOpen
  \bibfield  {author} {\bibinfo {author} {\bibfnamefont {J.~P.}\ \bibnamefont
  {Rodriguez}}, \bibinfo {author} {\bibfnamefont {M.~A.~N.}\ \bibnamefont
  {Araujo}}, \ and\ \bibinfo {author} {\bibfnamefont {P.~D.}\ \bibnamefont
  {Sacramento}},\ }\href {\doibase 10.1140/epjb/e2014-50198-9} {\bibfield
  {journal} {\bibinfo  {journal} {The European Physical Journal B}\ }\textbf
  {\bibinfo {volume} {87}},\ \bibinfo {pages} {163} (\bibinfo {year}
  {2014})}\BibitemShut {NoStop}%
\bibitem [{\citenamefont {Spa\l{}ek}\ \emph {et~al.}(2017)\citenamefont
  {Spa\l{}ek}, \citenamefont {Zegrodnik},\ and\ \citenamefont
  {Kaczmarczyk}}]{Spalek17}%
  \BibitemOpen
  \bibfield  {author} {\bibinfo {author} {\bibfnamefont {J.}~\bibnamefont
  {Spa\l{}ek}}, \bibinfo {author} {\bibfnamefont {M.}~\bibnamefont
  {Zegrodnik}}, \ and\ \bibinfo {author} {\bibfnamefont {J.}~\bibnamefont
  {Kaczmarczyk}},\ }\href {\doibase 10.1103/PhysRevB.95.024506} {\bibfield
  {journal} {\bibinfo  {journal} {Phys. Rev. B}\ }\textbf {\bibinfo {volume}
  {95}},\ \bibinfo {pages} {024506} (\bibinfo {year} {2017})}\BibitemShut
  {NoStop}%
\bibitem [{\citenamefont {Zegrodnik}\ and\ \citenamefont
  {Spa\l{}ek}(2017)}]{Zegrodnik17}%
  \BibitemOpen
  \bibfield  {author} {\bibinfo {author} {\bibfnamefont {M.}~\bibnamefont
  {Zegrodnik}}\ and\ \bibinfo {author} {\bibfnamefont {J.}~\bibnamefont
  {Spa\l{}ek}},\ }\href {\doibase 10.1103/PhysRevB.96.054511} {\bibfield
  {journal} {\bibinfo  {journal} {Phys. Rev. B}\ }\textbf {\bibinfo {volume}
  {96}},\ \bibinfo {pages} {054511} (\bibinfo {year} {2017})}\BibitemShut
  {NoStop}%
\bibitem [{\citenamefont {Ara\'ujo}\ \emph {et~al.}(2001)\citenamefont
  {Ara\'ujo}, \citenamefont {Peres},\ and\ \citenamefont
  {Sacramento}}]{Araujo01}%
  \BibitemOpen
  \bibfield  {author} {\bibinfo {author} {\bibfnamefont {M.~A.~N.}\
  \bibnamefont {Ara\'ujo}}, \bibinfo {author} {\bibfnamefont {N.~M.~R.}\
  \bibnamefont {Peres}}, \ and\ \bibinfo {author} {\bibfnamefont {P.~D.}\
  \bibnamefont {Sacramento}},\ }\href {\doibase 10.1103/PhysRevB.65.012503}
  {\bibfield  {journal} {\bibinfo  {journal} {Phys. Rev. B}\ }\textbf {\bibinfo
  {volume} {65}},\ \bibinfo {pages} {012503} (\bibinfo {year}
  {2001})}\BibitemShut {NoStop}%
\bibitem [{\citenamefont {Sacramento}(2003)}]{Sacramento03}%
  \BibitemOpen
  \bibfield  {author} {\bibinfo {author} {\bibfnamefont {P.~D.}\ \bibnamefont
  {Sacramento}},\ }\href {http://stacks.iop.org/0953-8984/15/i=36/a=315}
  {\bibfield  {journal} {\bibinfo  {journal} {Journal of Physics: Condensed
  Matter}\ }\textbf {\bibinfo {volume} {15}},\ \bibinfo {pages} {6285}
  (\bibinfo {year} {2003})}\BibitemShut {NoStop}%
\bibitem [{\citenamefont {Sacramento}\ \emph {et~al.}(2010)\citenamefont
  {Sacramento}, \citenamefont {Aparício},\ and\ \citenamefont
  {Nunes}}]{Sacramento10}%
  \BibitemOpen
  \bibfield  {author} {\bibinfo {author} {\bibfnamefont {P.~D.}\ \bibnamefont
  {Sacramento}}, \bibinfo {author} {\bibfnamefont {J.}~\bibnamefont
  {Aparício}}, \ and\ \bibinfo {author} {\bibfnamefont {G.~S.}\ \bibnamefont
  {Nunes}},\ }\href {http://stacks.iop.org/0953-8984/22/i=6/a=065702}
  {\bibfield  {journal} {\bibinfo  {journal} {Journal of Physics: Condensed
  Matter}\ }\textbf {\bibinfo {volume} {22}},\ \bibinfo {pages} {065702}
  (\bibinfo {year} {2010})}\BibitemShut {NoStop}%
\bibitem [{\citenamefont {Howczak}\ \emph {et~al.}(2013)\citenamefont
  {Howczak}, \citenamefont {Kaczmarczyk},\ and\ \citenamefont
  {Spa\l{}ek}}]{Howczak13}%
  \BibitemOpen
  \bibfield  {author} {\bibinfo {author} {\bibfnamefont {O.}~\bibnamefont
  {Howczak}}, \bibinfo {author} {\bibfnamefont {J.}~\bibnamefont
  {Kaczmarczyk}}, \ and\ \bibinfo {author} {\bibfnamefont {J.}~\bibnamefont
  {Spa\l{}ek}},\ }\href {\doibase 10.1002/pssb.201200774} {\bibfield  {journal}
  {\bibinfo  {journal} {physica status solidi (b)}\ }\textbf {\bibinfo {volume}
  {250}},\ \bibinfo {pages} {609} (\bibinfo {year} {2013})}\BibitemShut
  {NoStop}%
\bibitem [{\citenamefont {Wu}\ and\ \citenamefont {Tremblay}(2015)}]{Wu15}%
  \BibitemOpen
  \bibfield  {author} {\bibinfo {author} {\bibfnamefont {W.}~\bibnamefont
  {Wu}}\ and\ \bibinfo {author} {\bibfnamefont {A.-M.-S.}\ \bibnamefont
  {Tremblay}},\ }\href {\doibase 10.1103/PhysRevX.5.011019} {\bibfield
  {journal} {\bibinfo  {journal} {Phys. Rev. X}\ }\textbf {\bibinfo {volume}
  {5}},\ \bibinfo {pages} {011019} (\bibinfo {year} {2015})}\BibitemShut
  {NoStop}%
\bibitem [{\citenamefont {Wysoki\ifmmode~\acute{n}\else \'{n}\fi{}ski}\ \emph
  {et~al.}(2016)\citenamefont {Wysoki\ifmmode~\acute{n}\else \'{n}\fi{}ski},
  \citenamefont {Kaczmarczyk},\ and\ \citenamefont {Spa\l{}ek}}]{Wysokinski16}%
  \BibitemOpen
  \bibfield  {author} {\bibinfo {author} {\bibfnamefont {M.~M.}\ \bibnamefont
  {Wysoki\ifmmode~\acute{n}\else \'{n}\fi{}ski}}, \bibinfo {author}
  {\bibfnamefont {J.}~\bibnamefont {Kaczmarczyk}}, \ and\ \bibinfo {author}
  {\bibfnamefont {J.}~\bibnamefont {Spa\l{}ek}},\ }\href {\doibase
  10.1103/PhysRevB.94.024517} {\bibfield  {journal} {\bibinfo  {journal} {Phys.
  Rev. B}\ }\textbf {\bibinfo {volume} {94}},\ \bibinfo {pages} {024517}
  (\bibinfo {year} {2016})}\BibitemShut {NoStop}%
\bibitem [{\citenamefont {Xavier}\ and\ \citenamefont
  {Dagotto}(2008)}]{Xavier08}%
  \BibitemOpen
  \bibfield  {author} {\bibinfo {author} {\bibfnamefont {J.~C.}\ \bibnamefont
  {Xavier}}\ and\ \bibinfo {author} {\bibfnamefont {E.}~\bibnamefont
  {Dagotto}},\ }\href {\doibase 10.1103/PhysRevLett.100.146403} {\bibfield
  {journal} {\bibinfo  {journal} {Phys. Rev. Lett.}\ }\textbf {\bibinfo
  {volume} {100}},\ \bibinfo {pages} {146403} (\bibinfo {year}
  {2008})}\BibitemShut {NoStop}%
\bibitem [{\citenamefont {Liu}\ \emph {et~al.}(2012)\citenamefont {Liu},
  \citenamefont {Li}, \citenamefont {Zhang},\ and\ \citenamefont {Yu}}]{Yu12}%
  \BibitemOpen
  \bibfield  {author} {\bibinfo {author} {\bibfnamefont {Y.}~\bibnamefont
  {Liu}}, \bibinfo {author} {\bibfnamefont {H.}~\bibnamefont {Li}}, \bibinfo
  {author} {\bibfnamefont {G.-M.}\ \bibnamefont {Zhang}}, \ and\ \bibinfo
  {author} {\bibfnamefont {L.}~\bibnamefont {Yu}},\ }\href {\doibase
  10.1103/PhysRevB.86.024526} {\bibfield  {journal} {\bibinfo  {journal} {Phys.
  Rev. B}\ }\textbf {\bibinfo {volume} {86}},\ \bibinfo {pages} {024526}
  (\bibinfo {year} {2012})}\BibitemShut {NoStop}%
\bibitem [{\citenamefont {Bodensiek}\ \emph {et~al.}(2013)\citenamefont
  {Bodensiek}, \citenamefont {\ifmmode~\check{Z}\else \v{Z}\fi{}itko},
  \citenamefont {Vojta}, \citenamefont {Jarrell},\ and\ \citenamefont
  {Pruschke}}]{Bodensiek13}%
  \BibitemOpen
  \bibfield  {author} {\bibinfo {author} {\bibfnamefont {O.}~\bibnamefont
  {Bodensiek}}, \bibinfo {author} {\bibfnamefont {R.}~\bibnamefont
  {\ifmmode~\check{Z}\else \v{Z}\fi{}itko}}, \bibinfo {author} {\bibfnamefont
  {M.}~\bibnamefont {Vojta}}, \bibinfo {author} {\bibfnamefont
  {M.}~\bibnamefont {Jarrell}}, \ and\ \bibinfo {author} {\bibfnamefont
  {T.}~\bibnamefont {Pruschke}},\ }\href {\doibase
  10.1103/PhysRevLett.110.146406} {\bibfield  {journal} {\bibinfo  {journal}
  {Phys. Rev. Lett.}\ }\textbf {\bibinfo {volume} {110}},\ \bibinfo {pages}
  {146406} (\bibinfo {year} {2013})}\BibitemShut {NoStop}%
\bibitem [{\citenamefont {Asadzadeh}\ \emph {et~al.}(2014)\citenamefont
  {Asadzadeh}, \citenamefont {Fabrizio},\ and\ \citenamefont
  {Becca}}]{Asadzadeh14}%
  \BibitemOpen
  \bibfield  {author} {\bibinfo {author} {\bibfnamefont {M.~Z.}\ \bibnamefont
  {Asadzadeh}}, \bibinfo {author} {\bibfnamefont {M.}~\bibnamefont {Fabrizio}},
  \ and\ \bibinfo {author} {\bibfnamefont {F.}~\bibnamefont {Becca}},\ }\href
  {\doibase 10.1103/PhysRevB.90.205113} {\bibfield  {journal} {\bibinfo
  {journal} {Phys. Rev. B}\ }\textbf {\bibinfo {volume} {90}},\ \bibinfo
  {pages} {205113} (\bibinfo {year} {2014})}\BibitemShut {NoStop}%
\bibitem [{\citenamefont {Yu}\ \emph {et~al.}(2014)\citenamefont {Yu},
  \citenamefont {Guang-Ming},\ and\ \citenamefont {Lu}}]{Yu14}%
  \BibitemOpen
  \bibfield  {author} {\bibinfo {author} {\bibfnamefont {L.}~\bibnamefont
  {Yu}}, \bibinfo {author} {\bibfnamefont {Z.}~\bibnamefont {Guang-Ming}}, \
  and\ \bibinfo {author} {\bibfnamefont {Y.}~\bibnamefont {Lu}},\ }\href
  {http://stacks.iop.org/0256-307X/31/i=8/a=087102} {\bibfield  {journal}
  {\bibinfo  {journal} {Chinese Physics Letters}\ }\textbf {\bibinfo {volume}
  {31}},\ \bibinfo {pages} {087102} (\bibinfo {year} {2014})}\BibitemShut
  {NoStop}%
\bibitem [{\citenamefont {Lenz}\ \emph {et~al.}(2017)\citenamefont {Lenz},
  \citenamefont {Gezzi},\ and\ \citenamefont {Manmana}}]{Lenz17}%
  \BibitemOpen
  \bibfield  {author} {\bibinfo {author} {\bibfnamefont {B.}~\bibnamefont
  {Lenz}}, \bibinfo {author} {\bibfnamefont {R.}~\bibnamefont {Gezzi}}, \ and\
  \bibinfo {author} {\bibfnamefont {S.~R.}\ \bibnamefont {Manmana}},\ }\href
  {\doibase 10.1103/PhysRevB.96.155119} {\bibfield  {journal} {\bibinfo
  {journal} {Phys. Rev. B}\ }\textbf {\bibinfo {volume} {96}},\ \bibinfo
  {pages} {155119} (\bibinfo {year} {2017})}\BibitemShut {NoStop}%
\bibitem [{\citenamefont {Micnas}\ \emph {et~al.}(1990)\citenamefont {Micnas},
  \citenamefont {Ranninger},\ and\ \citenamefont {Robaszkiewicz}}]{Micnas1990}%
  \BibitemOpen
  \bibfield  {author} {\bibinfo {author} {\bibfnamefont {R.}~\bibnamefont
  {Micnas}}, \bibinfo {author} {\bibfnamefont {J.}~\bibnamefont {Ranninger}}, \
  and\ \bibinfo {author} {\bibfnamefont {S.}~\bibnamefont {Robaszkiewicz}},\
  }\href {\doibase 10.1103/RevModPhys.62.113} {\bibfield  {journal} {\bibinfo
  {journal} {Rev. Mod. Phys.}\ }\textbf {\bibinfo {volume} {62}},\ \bibinfo
  {pages} {113} (\bibinfo {year} {1990})}\BibitemShut {NoStop}%
\bibitem [{\citenamefont {Hewson}(1993)}]{Hewson1993}%
  \BibitemOpen
  \bibfield  {author} {\bibinfo {author} {\bibfnamefont {A.~C.}\ \bibnamefont
  {Hewson}},\ }\href {\doibase 10.1017/CBO9780511470752} {\emph {\bibinfo
  {title} {The Kondo Problem to Heavy Fermions}}},\ Cambridge Studies in
  Magnetism\ (\bibinfo  {publisher} {Cambridge University Press},\ \bibinfo
  {year} {1993})\BibitemShut {NoStop}%
\bibitem [{\citenamefont {Bertussi}\ \emph {et~al.}(2009)\citenamefont
  {Bertussi}, \citenamefont {Malvezzi}, \citenamefont {Paiva},\ and\
  \citenamefont {dos Santos}}]{Bertussi09}%
  \BibitemOpen
  \bibfield  {author} {\bibinfo {author} {\bibfnamefont {P.~R.}\ \bibnamefont
  {Bertussi}}, \bibinfo {author} {\bibfnamefont {A.~L.}\ \bibnamefont
  {Malvezzi}}, \bibinfo {author} {\bibfnamefont {T.}~\bibnamefont {Paiva}}, \
  and\ \bibinfo {author} {\bibfnamefont {R.~R.}\ \bibnamefont {dos Santos}},\
  }\href {\doibase 10.1103/PhysRevB.79.220513} {\bibfield  {journal} {\bibinfo
  {journal} {Phys. Rev. B}\ }\textbf {\bibinfo {volume} {79}},\ \bibinfo
  {pages} {220513} (\bibinfo {year} {2009})}\BibitemShut {NoStop}%
\bibitem [{\citenamefont {Karmakar}\ and\ \citenamefont
  {Majumdar}(2016)}]{Karmakar16}%
  \BibitemOpen
  \bibfield  {author} {\bibinfo {author} {\bibfnamefont {M.}~\bibnamefont
  {Karmakar}}\ and\ \bibinfo {author} {\bibfnamefont {P.}~\bibnamefont
  {Majumdar}},\ }\href {\doibase 10.1103/PhysRevB.93.195147} {\bibfield
  {journal} {\bibinfo  {journal} {Phys. Rev. B}\ }\textbf {\bibinfo {volume}
  {93}},\ \bibinfo {pages} {195147} (\bibinfo {year} {2016})}\BibitemShut
  {NoStop}%
\bibitem [{\citenamefont {Lynn}\ \emph {et~al.}(1997)\citenamefont {Lynn},
  \citenamefont {Skanthakumar}, \citenamefont {Huang}, \citenamefont {Sinha},
  \citenamefont {Hossain}, \citenamefont {Gupta}, \citenamefont {Nagarajan},\
  and\ \citenamefont {Godart}}]{Lynn1997}%
  \BibitemOpen
  \bibfield  {author} {\bibinfo {author} {\bibfnamefont {J.~W.}\ \bibnamefont
  {Lynn}}, \bibinfo {author} {\bibfnamefont {S.}~\bibnamefont {Skanthakumar}},
  \bibinfo {author} {\bibfnamefont {Q.}~\bibnamefont {Huang}}, \bibinfo
  {author} {\bibfnamefont {S.~K.}\ \bibnamefont {Sinha}}, \bibinfo {author}
  {\bibfnamefont {Z.}~\bibnamefont {Hossain}}, \bibinfo {author} {\bibfnamefont
  {L.~C.}\ \bibnamefont {Gupta}}, \bibinfo {author} {\bibfnamefont
  {R.}~\bibnamefont {Nagarajan}}, \ and\ \bibinfo {author} {\bibfnamefont
  {C.}~\bibnamefont {Godart}},\ }\href {\doibase 10.1103/PhysRevB.55.6584}
  {\bibfield  {journal} {\bibinfo  {journal} {Phys. Rev. B}\ }\textbf {\bibinfo
  {volume} {55}},\ \bibinfo {pages} {6584} (\bibinfo {year}
  {1997})}\BibitemShut {NoStop}%
\bibitem [{\citenamefont {Schmidt}\ and\ \citenamefont
  {Braun}(1997)}]{Schmidt1997}%
  \BibitemOpen
  \bibfield  {author} {\bibinfo {author} {\bibfnamefont {H.}~\bibnamefont
  {Schmidt}}\ and\ \bibinfo {author} {\bibfnamefont {H.~F.}\ \bibnamefont
  {Braun}},\ }\href {\doibase 10.1103/PhysRevB.55.8497} {\bibfield  {journal}
  {\bibinfo  {journal} {Phys. Rev. B}\ }\textbf {\bibinfo {volume} {55}},\
  \bibinfo {pages} {8497} (\bibinfo {year} {1997})}\BibitemShut {NoStop}%
\bibitem [{\citenamefont {\surname{ElMassalami}}\ \emph
  {et~al.}(2012)\citenamefont {\surname{ElMassalami}}, \citenamefont {Takeya},
  \citenamefont {Ouladdiaf}, \citenamefont {Maia~Filho}, \citenamefont {Gomes},
  \citenamefont {Paiva},\ and\ \citenamefont {dos Santos}}]{ElMassalami2012}%
  \BibitemOpen
  \bibfield  {author} {\bibinfo {author} {\bibfnamefont {M.}~\bibnamefont
  {\surname{ElMassalami}}}, \bibinfo {author} {\bibfnamefont {H.}~\bibnamefont
  {Takeya}}, \bibinfo {author} {\bibfnamefont {B.}~\bibnamefont {Ouladdiaf}},
  \bibinfo {author} {\bibfnamefont {R.}~\bibnamefont {Maia~Filho}}, \bibinfo
  {author} {\bibfnamefont {A.~M.}\ \bibnamefont {Gomes}}, \bibinfo {author}
  {\bibfnamefont {T.}~\bibnamefont {Paiva}}, \ and\ \bibinfo {author}
  {\bibfnamefont {R.~R.}\ \bibnamefont {dos Santos}},\ }\href {\doibase
  10.1103/PhysRevB.85.174412} {\bibfield  {journal} {\bibinfo  {journal} {Phys.
  Rev. B}\ }\textbf {\bibinfo {volume} {85}},\ \bibinfo {pages} {174412}
  (\bibinfo {year} {2012})}\BibitemShut {NoStop}%
\bibitem [{\citenamefont {\surname{ElMassalami}}\ \emph
  {et~al.}(2013)\citenamefont {\surname{ElMassalami}}, \citenamefont {Gomes},
  \citenamefont {Paiva}, \citenamefont {\surname{dos Santos}},\ and\
  \citenamefont {Takeya}}]{ElMassalami2013}%
  \BibitemOpen
  \bibfield  {author} {\bibinfo {author} {\bibfnamefont {M.}~\bibnamefont
  {\surname{ElMassalami}}}, \bibinfo {author} {\bibfnamefont {A.~M.}\
  \bibnamefont {Gomes}}, \bibinfo {author} {\bibfnamefont {T.}~\bibnamefont
  {Paiva}}, \bibinfo {author} {\bibfnamefont {R.~R.}\ \bibnamefont
  {\surname{dos Santos}}}, \ and\ \bibinfo {author} {\bibfnamefont
  {H.}~\bibnamefont {Takeya}},\ }\href {\doibase
  http://dx.doi.org/10.1016/j.jmmm.2013.01.044} {\bibfield  {journal} {\bibinfo
   {journal} {J. Magn. Magn. Mater.}\ }\textbf {\bibinfo {volume} {335}},\
  \bibinfo {pages} {163 } (\bibinfo {year} {2013})}\BibitemShut {NoStop}%
\bibitem [{\citenamefont {\surname{ElMassalami}}\ \emph
  {et~al.}(2014)\citenamefont {\surname{ElMassalami}}, \citenamefont {Takeya},
  \citenamefont {Ouladdiaf}, \citenamefont {Gomes}, \citenamefont {Paiva},\
  and\ \citenamefont {\surname{dos Santos}}}]{ElMassalami2014}%
  \BibitemOpen
  \bibfield  {author} {\bibinfo {author} {\bibfnamefont {M.}~\bibnamefont
  {\surname{ElMassalami}}}, \bibinfo {author} {\bibfnamefont {H.}~\bibnamefont
  {Takeya}}, \bibinfo {author} {\bibfnamefont {B.}~\bibnamefont {Ouladdiaf}},
  \bibinfo {author} {\bibfnamefont {A.~M.}\ \bibnamefont {Gomes}}, \bibinfo
  {author} {\bibfnamefont {T.}~\bibnamefont {Paiva}}, \ and\ \bibinfo {author}
  {\bibfnamefont {R.~R.}\ \bibnamefont {\surname{dos Santos}}},\ }\href
  {\doibase http://dx.doi.org/10.1016/j.jmmm.2014.07.026} {\bibfield  {journal}
  {\bibinfo  {journal} {J. Magn. Magn. Mater.}\ }\textbf {\bibinfo {volume}
  {372}},\ \bibinfo {pages} {74 } (\bibinfo {year} {2014})}\BibitemShut
  {NoStop}%
\bibitem [{\citenamefont {Costa}\ \emph {et~al.}(2017)\citenamefont {Costa},
  \citenamefont {\surname{Pimentel de Lima}},\ and\ \citenamefont {\surname{dos
  Santos}}}]{Costa17a}%
  \BibitemOpen
  \bibfield  {author} {\bibinfo {author} {\bibfnamefont {N.~C.}\ \bibnamefont
  {Costa}}, \bibinfo {author} {\bibfnamefont {J.}~\bibnamefont
  {\surname{Pimentel de Lima}}}, \ and\ \bibinfo {author} {\bibfnamefont
  {R.~R.}\ \bibnamefont {\surname{dos Santos}}},\ }\href {\doibase
  http://dx.doi.org/10.1016/j.jmmm.2016.09.061} {\bibfield  {journal} {\bibinfo
   {journal} {J. Magn. Magn. Mater.}\ }\textbf {\bibinfo {volume} {423}},\
  \bibinfo {pages} {74} (\bibinfo {year} {2017})}\BibitemShut {NoStop}%
\bibitem [{\citenamefont {Paiva}\ \emph {et~al.}(2004)\citenamefont {Paiva},
  \citenamefont {dos Santos}, \citenamefont {Scalettar},\ and\ \citenamefont
  {Denteneer}}]{Paiva2004}%
  \BibitemOpen
  \bibfield  {author} {\bibinfo {author} {\bibfnamefont {T.}~\bibnamefont
  {Paiva}}, \bibinfo {author} {\bibfnamefont {R.~R.}\ \bibnamefont {dos
  Santos}}, \bibinfo {author} {\bibfnamefont {R.~T.}\ \bibnamefont
  {Scalettar}}, \ and\ \bibinfo {author} {\bibfnamefont {P.~J.~H.}\
  \bibnamefont {Denteneer}},\ }\href {\doibase 10.1103/PhysRevB.69.184501}
  {\bibfield  {journal} {\bibinfo  {journal} {Phys. Rev. B}\ }\textbf {\bibinfo
  {volume} {69}},\ \bibinfo {pages} {184501} (\bibinfo {year}
  {2004})}\BibitemShut {NoStop}%
\bibitem [{\citenamefont {Spa\l{}ek}(1988)}]{Spalek88}%
  \BibitemOpen
  \bibfield  {author} {\bibinfo {author} {\bibfnamefont {J.}~\bibnamefont
  {Spa\l{}ek}},\ }\href {\doibase 10.1103/PhysRevB.38.208} {\bibfield
  {journal} {\bibinfo  {journal} {Phys. Rev. B}\ }\textbf {\bibinfo {volume}
  {38}},\ \bibinfo {pages} {208} (\bibinfo {year} {1988})}\BibitemShut
  {NoStop}%
\bibitem [{\citenamefont {Cyrot}(1986)}]{Cyrot86}%
  \BibitemOpen
  \bibfield  {author} {\bibinfo {author} {\bibfnamefont {M.}~\bibnamefont
  {Cyrot}},\ }\href {\doibase https://doi.org/10.1016/0038-1098(86)90458-8}
  {\bibfield  {journal} {\bibinfo  {journal} {Solid State Communications}\
  }\textbf {\bibinfo {volume} {60}},\ \bibinfo {pages} {253 } (\bibinfo {year}
  {1986})}\BibitemShut {NoStop}%
\bibitem [{\citenamefont {Gehring}\ and\ \citenamefont
  {Major}(1994)}]{Gehring94}%
  \BibitemOpen
  \bibfield  {author} {\bibinfo {author} {\bibfnamefont {G.~A.}\ \bibnamefont
  {Gehring}}\ and\ \bibinfo {author} {\bibfnamefont {L.~E.}\ \bibnamefont
  {Major}},\ }\href {http://stacks.iop.org/0953-8984/6/i=2/a=020} {\bibfield
  {journal} {\bibinfo  {journal} {Journal of Physics: Condensed Matter}\
  }\textbf {\bibinfo {volume} {6}},\ \bibinfo {pages} {495} (\bibinfo {year}
  {1994})}\BibitemShut {NoStop}%
\bibitem [{\citenamefont {Masuda}\ and\ \citenamefont
  {Yamamoto}(2013)}]{Masuda13}%
  \BibitemOpen
  \bibfield  {author} {\bibinfo {author} {\bibfnamefont {K.}~\bibnamefont
  {Masuda}}\ and\ \bibinfo {author} {\bibfnamefont {D.}~\bibnamefont
  {Yamamoto}},\ }\href {\doibase 10.1103/PhysRevB.87.014516} {\bibfield
  {journal} {\bibinfo  {journal} {Phys. Rev. B}\ }\textbf {\bibinfo {volume}
  {87}},\ \bibinfo {pages} {014516} (\bibinfo {year} {2013})}\BibitemShut
  {NoStop}%
\bibitem [{\citenamefont {Masuda}\ and\ \citenamefont
  {Yamamoto}(2015)}]{Masuda15}%
  \BibitemOpen
  \bibfield  {author} {\bibinfo {author} {\bibfnamefont {K.}~\bibnamefont
  {Masuda}}\ and\ \bibinfo {author} {\bibfnamefont {D.}~\bibnamefont
  {Yamamoto}},\ }\href {\doibase 10.1103/PhysRevB.91.104508} {\bibfield
  {journal} {\bibinfo  {journal} {Phys. Rev. B}\ }\textbf {\bibinfo {volume}
  {91}},\ \bibinfo {pages} {104508} (\bibinfo {year} {2015})}\BibitemShut
  {NoStop}%
\bibitem [{\citenamefont {Lacroix}\ and\ \citenamefont
  {Cyrot}(1979)}]{Lacroix1979}%
  \BibitemOpen
  \bibfield  {author} {\bibinfo {author} {\bibfnamefont {C.}~\bibnamefont
  {Lacroix}}\ and\ \bibinfo {author} {\bibfnamefont {M.}~\bibnamefont
  {Cyrot}},\ }\href {\doibase 10.1103/PhysRevB.20.1969} {\bibfield  {journal}
  {\bibinfo  {journal} {Phys. Rev. B}\ }\textbf {\bibinfo {volume} {20}},\
  \bibinfo {pages} {1969} (\bibinfo {year} {1979})}\BibitemShut {NoStop}%
\bibitem [{\citenamefont {\surname{dos Santos}}(1994)}]{dosSantos94b}%
  \BibitemOpen
  \bibfield  {author} {\bibinfo {author} {\bibfnamefont {R.~R.}\ \bibnamefont
  {\surname{dos Santos}}},\ }\href {\doibase 10.1103/PhysRevB.50.635}
  {\bibfield  {journal} {\bibinfo  {journal} {Phys. Rev. B}\ }\textbf {\bibinfo
  {volume} {50}},\ \bibinfo {pages} {635} (\bibinfo {year} {1994})}\BibitemShut
  {NoStop}%
\bibitem [{Note1()}]{Note1}%
  \BibitemOpen
  \bibinfo {note} {\label {foot25} The meaning of \protect \emph {weakly
  superconducting} in the context of this paper has no connection with the more
  common usage of the term, denoting the type of superconductivity occurring on
  a surface or a grain boundary; here it simply denotes a weakening of the
  gap.}\BibitemShut {Stop}%
\bibitem [{Note2()}]{Note2}%
  \BibitemOpen
  \bibinfo {note} {\label {foot21} We recall that the phase boundaries are
  determined by crossings in the lowest ground state energy, and they coincide
  with those obtained by the vanishing of the different order parameters; when
  approaching the spiral magnetic phase at small $U/t$, we observe $\Delta $
  being smaller than a given tolerance, in accordance with the decreasing
  exponential behaviour.}\BibitemShut {Stop}%
\bibitem [{\citenamefont {Peters}\ and\ \citenamefont
  {Kawakami}(2015)}]{Peters15}%
  \BibitemOpen
  \bibfield  {author} {\bibinfo {author} {\bibfnamefont {R.}~\bibnamefont
  {Peters}}\ and\ \bibinfo {author} {\bibfnamefont {N.}~\bibnamefont
  {Kawakami}},\ }\href {\doibase 10.1103/PhysRevB.92.075103} {\bibfield
  {journal} {\bibinfo  {journal} {Phys. Rev. B}\ }\textbf {\bibinfo {volume}
  {92}},\ \bibinfo {pages} {075103} (\bibinfo {year} {2015})}\BibitemShut
  {NoStop}%
\bibitem [{\citenamefont {Peters}\ and\ \citenamefont
  {Kawakami}(2017)}]{Peters17}%
  \BibitemOpen
  \bibfield  {author} {\bibinfo {author} {\bibfnamefont {R.}~\bibnamefont
  {Peters}}\ and\ \bibinfo {author} {\bibfnamefont {N.}~\bibnamefont
  {Kawakami}},\ }\href {\doibase 10.1103/PhysRevB.96.115158} {\bibfield
  {journal} {\bibinfo  {journal} {Phys. Rev. B}\ }\textbf {\bibinfo {volume}
  {96}},\ \bibinfo {pages} {115158} (\bibinfo {year} {2017})}\BibitemShut
  {NoStop}%
\bibitem [{\citenamefont {Lanat\`a}\ \emph {et~al.}(2008)\citenamefont
  {Lanat\`a}, \citenamefont {Barone},\ and\ \citenamefont
  {Fabrizio}}]{Lanata08}%
  \BibitemOpen
  \bibfield  {author} {\bibinfo {author} {\bibfnamefont {N.}~\bibnamefont
  {Lanat\`a}}, \bibinfo {author} {\bibfnamefont {P.}~\bibnamefont {Barone}}, \
  and\ \bibinfo {author} {\bibfnamefont {M.}~\bibnamefont {Fabrizio}},\ }\href
  {\doibase 10.1103/PhysRevB.78.155127} {\bibfield  {journal} {\bibinfo
  {journal} {Phys. Rev. B}\ }\textbf {\bibinfo {volume} {78}},\ \bibinfo
  {pages} {155127} (\bibinfo {year} {2008})}\BibitemShut {NoStop}%
\bibitem [{\citenamefont {Watanabe}\ and\ \citenamefont
  {Ogata}(2007)}]{Watanabe07}%
  \BibitemOpen
  \bibfield  {author} {\bibinfo {author} {\bibfnamefont {H.}~\bibnamefont
  {Watanabe}}\ and\ \bibinfo {author} {\bibfnamefont {M.}~\bibnamefont
  {Ogata}},\ }\href {\doibase 10.1103/PhysRevLett.99.136401} {\bibfield
  {journal} {\bibinfo  {journal} {Phys. Rev. Lett.}\ }\textbf {\bibinfo
  {volume} {99}},\ \bibinfo {pages} {136401} (\bibinfo {year}
  {2007})}\BibitemShut {NoStop}%
\bibitem [{\citenamefont {Asadzadeh}\ \emph {et~al.}(2013)\citenamefont
  {Asadzadeh}, \citenamefont {Becca},\ and\ \citenamefont
  {Fabrizio}}]{Asadzadeh13}%
  \BibitemOpen
  \bibfield  {author} {\bibinfo {author} {\bibfnamefont {M.~Z.}\ \bibnamefont
  {Asadzadeh}}, \bibinfo {author} {\bibfnamefont {F.}~\bibnamefont {Becca}}, \
  and\ \bibinfo {author} {\bibfnamefont {M.}~\bibnamefont {Fabrizio}},\ }\href
  {\doibase 10.1103/PhysRevB.87.205144} {\bibfield  {journal} {\bibinfo
  {journal} {Phys. Rev. B}\ }\textbf {\bibinfo {volume} {87}},\ \bibinfo
  {pages} {205144} (\bibinfo {year} {2013})}\BibitemShut {NoStop}%
\bibitem [{Note3()}]{Note3}%
  \BibitemOpen
  \bibinfo {note} {Here we use the term N\'eel temperature, $T_N$, in a broader
  sense, encompassing the critical temperature for spiral states as
  well.}\BibitemShut {Stop}%
\bibitem [{\citenamefont {Doniach}(1977)}]{Doniach1977}%
  \BibitemOpen
  \bibfield  {author} {\bibinfo {author} {\bibfnamefont {S.}~\bibnamefont
  {Doniach}},\ }\href {\doibase http://dx.doi.org/10.1016/0378-4363(77)90190-5}
  {\bibfield  {journal} {\bibinfo  {journal} {Physica B+C}\ }\textbf {\bibinfo
  {volume} {91}},\ \bibinfo {pages} {231 } (\bibinfo {year}
  {1977})}\BibitemShut {NoStop}%
\end{thebibliography}%

\end{document}